\documentclass[aps,prd,twocolumn,groupedaddress,floatfix]{revtex4}
\usepackage{amsmath,graphicx}

\begin{document}


\title{Schwarzschild Tests of the Wahlquist-Estabrook-Buchman-Bardeen 
Tetrad Formulation for Numerical Relativity}


\author{L. T. Buchman}
\affiliation{Jet Propulsion Laboratory, California Institute of Technology, 
Pasadena, CA}
\author{J. M. Bardeen}
\affiliation{Physics Department, University of Washington, Seattle, WA}


\date{\today}

\begin{abstract}
A first order symmetric hyperbolic tetrad formulation 
of the Einstein equations developed by
Estabrook and Wahlquist and put into a form suitable for numerical
relativity by Buchman and Bardeen (the WEBB formulation) is adapted to
explicit spherical symmetry and tested for accuracy
and stability in the evolution of spherically symmetric black holes
(the Schwarzschild geometry).  The lapse and shift, 
which specify the evolution of the coordinates relative
to the tetrad congruence, are reset at frequent time intervals
to keep the constant-time hypersurfaces nearly
orthogonal to the tetrad congruence and the spatial coordinate
satisfying a kind of minimal rate of strain condition.  
By arranging through initial conditions that the constant-time 
hypersurfaces are asymptotically hyperbolic, we simplify
the boundary value problem and improve stability of the evolution.  
Results are obtained for both tetrad gauges 
(``Nester'' and ``Lorentz'') of the 
WEBB formalism using finite difference numerical methods.  
We are able to obtain stable unconstrained evolution with the Nester 
gauge for certain initial conditions, but not with the Lorentz gauge.

\end{abstract}


\maketitle

\setlength{\unitlength}{.15cm}

\section{Introduction}

An orthonormal tetrad approach to numerical relativity has several attractive 
features.  The metric is trivial, and most of the variables are coordinate 
scalars, which eliminates derivatives of the shift vector from most of the
equations.  There are only twenty-four connection coefficient variables in general, 
the Ricci rotation coefficients, as opposed to the forty connection coefficients
in a metric-based formulation.  While one does have to evolve the tetrad 
vectors in place of the metric, one does not have to deal with nonlinearities
in the equations associated with the inverse metric.
A number of tetrad and triad formulations for general relativity
have been proposed \cite{M61,newman62,EW64,AA86,AA87,HF96,VPE96,
ERW97,VEU97,ILR98,YS99,YS00,Jantzen01,CBY02,BuBa03,Frauendiener04,Bardeen05}, 
but they have not been as widely used in
numerical relativity as standard 3+1 metric-based formulations.  This paper 
describes the numerical implementation and tests of one such scheme, 
the WEBB formulation \cite{BuBa03}, in spherically 
symmetric vacuum black hole spacetimes.  
See Estabrook's papers \cite{Estabrook05a,Estabrook05b} 
for an in-depth analysis of the mathematical structure of these equations
(without specified gauge conditions).

The WEBB scheme takes as its primary variables the tetrad 
connection coefficients, and incorporates one of two alternative dynamic 
gauge conditions to evolve the acceleration and angular velocity of the tetrads,
either the Nester gauge \cite{JN92} or the Lorentz gauge \cite{VPE96}.
The evolution equations constitute a 
first-order symmetrizable hyperbolic system in which all variables propagate
either along the light cone or along the tetrad congruence, with the
tetrad gauge information propagating along the light cone.  Both the
Nester and Lorentz gauges do not in general 
preserve hypersurface orthogonality of the tetrad 
congruence.  The coordinate evolution is controlled by a tetrad lapse function
and shift vector which we do not evolve dynamically, but rather
reset periodically to keep the constant-time hypersurfaces 
nearly (but not exactly) orthogonal to the congruence worldlines, 
and to maintain a minimal deformation condition on the spatial 
coordinates which is similar to the minimal strain condition of
Smarr and York \cite{SY78} in the $3+1$ context.

Spherically symmetric spacetimes, while in a sense trivial in that 
they do not allow any gravitational waves to be present, 
provide the challenge of 
maintaining a stable numerical evolution in the presence of an event horizon. 
Additionally, they require dealing with both an excision inner boundary 
and an outer boundary.  In our code, the numerical grid extends from just
inside the event horizon to around $R=20~M$, where $R$ is the circumferential
radius of a two-sphere.  Our initial slices are constructed 
so that the congruence worldlines point out of the grid at
both boundaries.  This forces all of the eigenmodes to propagate
out of the grid at the inner boundary, and the eigenmodes
which travel along the congruence worldlines, as well those
which travel along the outgoing light cones, to propagate out of the
grid at the outer boundary.  Even so, there are two eigenmodes 
(a ``constraint'' mode and a gauge mode) traveling 
at the speed of light into the grid at the outer boundary.
Boundary conditions for the ``constraint'' mode are determined
according to constraint-preserving boundary conditions 
\cite{Stewart98,FN99,IR02,BaBu02,SW03,CPRST03,SSW02,CLT02,CS03,
frittelligomez04,LSKPST04,KLSBP05,ST05},
which insure that the information entering the numerical grid is consistent
with the constraint equations.  There is no physical constraint on the incoming
gauge mode; we choose to keep its amplitude fixed as set in the 
initial conditions.

This paper is one of a relatively small number of numerical tests of
tetrad/triad formulations in vacuum general relativity
(see \cite{VP97,Frauendiener98,SY00,YS01,DF04,Frauendiener04}).  
We find that it is possible to
achieve reasonable long term stability evolving the spherically symmetric 
Schwarzschild geometry using the WEBB equations with the Nester gauge,
but only for rather special initial conditions.  Many other equation 
formulations and tetrad gauge conditions are possible in the context of
orthonormal frames and remain to be explored.  Tests in the limited context 
of spherical symmetry and one-dimension are far from sufficient to 
establish the viability of a particular formulation, but can serve to establish
lack of viability.

\section{Variables}

The application of the three-dimensional WEBB formalism 
presented in \cite{BuBa03} 
to one-dimensional Schwarzschild black holes requires
the construction of an orthonormal tetrad field well-behaved everywhere
outside and in the vicinity of the event horizon.  The most obvious
basis vectors for a spherically symmetric spacetime consist of a timelike 
vector field $\boldsymbol{e}_0$ and a spacelike vector field 
$\boldsymbol{e}_{\hat{r}}$, both orthogonal to the two-spheres generated 
by the symmetry, and two spacelike vector fields tangent to the two-spheres.  
The timelike vector field  defines a 
timelike congruence of worldlines orthogonal to the two-spheres.  The 
problem is that orthonormal vector fields tangent to a two-sphere cannot 
continuously be defined everywhere on the two-sphere.  If these tangent 
vectors are chosen to be aligned with angular polar coordinates, 
$\boldsymbol{e}_{\hat{\theta}}$ and $\boldsymbol{e}_{\hat{\phi}}$, 
they are degenerate at the poles, $\theta = 0$ and $\theta = \pi$.  
These ``spherical'' spacelike basis vectors are unsuitable for direct
use as the tetrad vectors in the WEBB tetrad formalism.  There would be
singularities in some of the connection coefficients (Ricci rotation 
coefficients) at the poles.  Instead, we must define a ``Cartesian'' triad 
of spacelike vectors which is rotated from the spherical triad.  A simple 
way to do this is to invoke the same rotation as a function of the polar 
angles that takes spherical to Cartesian basis vectors in flat space, namely:
\begin{equation}
\label{e1}
\boldsymbol{e}_1=\cos{\phi}~\sin{\theta}~\boldsymbol{e}_{\hat{r}}+
\cos{\phi}~\cos{\theta}~\boldsymbol{e}_{\hat{\theta}}-
\sin{\phi}~\boldsymbol{e}_{\hat{\phi}},
\end{equation}
\begin{equation}
\label{e2}
\boldsymbol{e}_2=\sin{\phi}~\sin{\theta}~\boldsymbol{e}_{\hat{r}}+
\sin{\phi}~\cos{\theta}~\boldsymbol{e}_{\hat{\theta}}+
\cos{\phi}~\boldsymbol{e}_{\hat{\phi}},
\end{equation}
\begin{equation}
\label{e3}
\boldsymbol{e}_3=\cos{\theta}~\boldsymbol{e}_{\hat{r}}-
\sin{\theta}~\boldsymbol{e}_{\hat{\theta}}.
\end{equation}
While the Cartesian tetrad must be used to define the WEBB variables, once
these are defined it is convenient to rotate back to the spherical triad 
to take explicit advantage of the spherical symmetry.

As in any Cauchy formulation for numerical relativity, the evolution
of the spacetime is described by a sequence of spacelike hypersurfaces.  
Since the state of the system is specified on such a constant-time 
hypersurface, spatial derivatives must be evaluated at constant time.  
However, the spatial triad vectors are not in general tangent to the
constant-time hypersurface.
As described in \cite{BuBa03}, we decompose the spatial triad vectors 
into a timelike component parallel to the congruence 
and a spacelike component tangent to the hypersurface:
\begin{equation}
\boldsymbol{e}_a=A_a~\boldsymbol{e}_0+B_a^{~k}~\frac{\partial}{\partial x^k}.
\end{equation}
The three-vector $\boldsymbol{B}_a$ is not a unit vector if $A_a$ is not zero.
The vectors $\boldsymbol{e}_{\hat{\theta}}$ and $\boldsymbol{e}_{\hat{\phi}}$
are tangent to the hypersurfaces, so $A_{\hat{\theta}}=A_{\hat{\phi}}=0$. 
Thus, $A_a$ only has one degree of freedom, in the $\boldsymbol{e}_{\hat{r}}$
direction.  $A_{\hat{r}}$ is the radial 3-velocity of a tetrad observer
with respect to an observer at rest in the constant-time hypersurface. 
We can now write the spherical triad vectors as 
\begin{equation}
\boldsymbol{e}_{\hat{r}}=A_{\hat{r}}~\boldsymbol{e}_0+B_{\hat{r}}^{~r}
~~{\partial_r},
~~~\boldsymbol{e}_{\hat{\theta}}=B_{\hat{\theta}}^{~\theta}~~
{\partial_{\theta}},
~~~\boldsymbol{e}_{\hat{\phi}}=B_{\hat{\phi}}^{~\phi}~~{\partial_{\phi}},
\end{equation}
where $r$ is the radial coordinate in the hypersurface.  
Since $\boldsymbol{B}_{\hat{\theta}}$ and $\boldsymbol{B}_{\hat{\phi}}$ 
are unit vectors, they can be found from the metric.  We find it convenient 
to define new symbols such that 
\begin{equation}
\label{Bcomps}
B_{\hat{r}}^{~r} \equiv B_R = e^{-{\lambda}},~~~
B_{\hat{\theta}}^{~\theta} \equiv B_T=\frac{1}{R},~~~
B_{\hat{\phi}}^{~\phi}=\frac{B_T}{\sin{\theta}},
\end{equation}
with $R$ the circumferential radius of the two-sphere.

The directional derivative along the timelike vector of the tetrad can be 
related to coordinate derivatives by defining a tetrad ``lapse'' $\alpha$ 
and ``shift'' vector $\beta^k$.  The shift has only a radial component,
so
\begin{equation}
\label{congruencedir}
D_0=\frac{1}{\alpha}~(\partial_t-\beta^r~\partial_r).
\end{equation}
Unless the tetrad congruence is orthogonal to the constant-time hypersurfaces, 
the tetrad lapse and shift are different from the standard 3+1 lapse and 
shift.

The directional derivatives
along the spatial triad directions are 
\begin{equation}
D_a=e_a^{~\mu}~\frac{\partial}{\partial x^{\mu}}
\end{equation}
for the Cartesian triad, which is related to the spherical triad by 
Eqs. (\ref{e1}),  (\ref{e2}), and (\ref{e3}).
The spherical directional derivatives are 
\begin{equation}
\label{rhatdir}
D_{\hat{r}}=A_{\hat{r}}~D_0+e^{-\lambda}~\partial_r,
\end{equation}
\begin{equation}
\label{transvdir}
D_{\hat{\theta}}=\frac{1}{R}~\partial_{\theta},
~~D_{\hat{\phi}}=\frac{1}{R~\sin{\theta}}~\partial_{\phi}.
\end{equation}

The twenty-four connection coefficients are derived
from the commutators of the Cartesian tetrad directional derivatives.
In the WEBB formulation, a space-time split is made which groups the 
connection coefficients into two $3 \times 3$ dyadic matrices 
$N_{ab}\equiv\frac{1}{2}\:\varepsilon_{bcd}\:\Gamma_{cda}$ and 
$K_{ab}\equiv\Gamma_{b0a}$, plus two spatial vectors, the 
acceleration $a_b=\Gamma_{b00}$ and the angular velocity 
(relative to Fermi-Walker transport) $\omega_b=\frac{1}{2}
\:\varepsilon_{bcd}\:\Gamma_{dc0}$ of the tetrad frames.

As a consequence of the underlying spherical symmetry,
only the anti-symmetric part of the  $N_{ab}$ is non-zero.
The vector $\boldsymbol{n}$ with Cartesian components
$n_{a}\equiv\frac{1}{2}\:\varepsilon_{abc}\:N_{bc}$ points in the radial
direction, $\boldsymbol{n}=n_{\hat{r}}~\boldsymbol{e}_{\hat{r}}$,
and the constraint equations arising from commutators of the spatial 
tetrad vectors reduce to
\begin{equation}
\label{Rprimeconstraint}
n_{\hat{r}}=~\frac{1-D_{\hat{r}}~R}{R}.
\end{equation}

The spherical symmetry also means that the  $K_{ab}$ are 
symmetric, which is the condition that the tetrad congruence has zero 
vorticity and is orthogonal to some set of spacelike hypersurfaces.  
The $K_{ab}$ are the Cartesian components of the extrinsic 
curvature tensor of these hypersurfaces (not necessarily the same as 
constant-time hypersurfaces).  There are only two independent 
degrees of freedom in the extrinsic curvature, since the only 
non-zero components of the extrinsic curvature with respect to the
spherical triad basis are 
\begin{equation}
\label{spherK}
K_{\hat{r}\hat{r}} \equiv K_R,~~ K_{\hat{\theta}\hat{\theta}} =
K_{\hat{\phi}\hat{\phi}} \equiv K_T.
\end{equation}

Finally, the angular velocity $\omega_b$ of a spherically symmetric 
congruence is identically zero, and the acceleration can only point in 
the radial direction perpendicular to the congruence worldlines, so the 
only non-zero component relative to the
spherical triad basis is $a_{\hat{r}}$.  The calculation from the 
Cartesian basis gives
\begin{equation}
a_1=\cos{\phi}\sin{\theta}~a_{\hat{r}},~
a_2=\sin{\phi}\sin{\theta}~a_{\hat{r}},~
a_3=\cos{\theta}~a_{\hat{r}}.
\end{equation}

The WEBB equations are greatly simplified if
$K_R$, $K_T$, $n_{\hat{r}}$, $a_{\hat{r}}$,
$B_R$, $B_T$, and $A_{\hat{r}}$ as used as variables instead of the 
Cartesian components.
This results in a reduction of the total
number of variables from thirty-six to seven.  The lapse function and shift
vector are not evolved dynamically, but are reset periodically to optimize 
the evolution of the coordinates.

The spacetime metric is obtained
by first calculating $g^{\mu\nu}=\eta^{\alpha\beta}\:e_{\alpha}^{~\mu}
\:e_{\beta}^{~\nu}$.  The $g^{\mu\nu}$ matrix is then inverted to give
$g_{\mu\nu}$.  The resulting metric is:
\begin{multline}
\label{spacetimemetric}
ds^2=[-\alpha^2+{\beta^r}^2~e^{2\lambda}~(1-A_{\hat{r}}^2)
+2~e^{\lambda}~\alpha~\beta^r~A_{\hat{r}}]~dt^2\\
+2~e^{\lambda}~[\alpha~A_{\hat{r}}+\beta^r~e^{\lambda}
~(1-A_{\hat{r}}^2)]~dr~dt~~~~~~~~~~~~~~~\\
+e^{2\lambda}~(1-A_{\hat{r}}^2)~dr^2+R^2~d\theta^2+R^2~\sin^2{\theta}~d\phi^2.
\end{multline}
The spatial metric of the constant-time hypersurface is
\begin{equation}
\label{metric}
dl^2=e^{2\lambda}~(1-A_{\hat{r}}^2)~dr^2
+R^2~d\theta^2+R^2~\sin^2{\theta}~d\phi^2.
\end{equation}

\section{Tetrad Quasievolution and Quasiconstraint Equations}
\label{sectetradeqn}

The true evolution equations must be equations for partial time derivatives 
of the variables in terms of partial spatial derivatives 
and source terms which 
are functions of the variables.  The natural result of the WEBB formalism
is equations relating directional derivatives.  Because in general there are
time derivatives hidden inside the spatial directional derivatives, we 
call the equations expressed in terms of the tetrad directional derivatives 
quasievolution equations 
(if they contain an explicit $D_0$) or quasiconstraint 
equations if they contain only spatial directional derivatives.  (See
the WEBB paper \cite{BuBa03} for a complete discussion.)  Because they are
simpler, we first discuss the quasievolution and quasiconstraint equations.

Since the Einstein equations are covariant under rotations of the tetrad,
the quasievolution equations for
$K_R,~K_T,$ and $n_{\hat{r}}$, and the
quasiconstraint equations for $K_T$ and $n_{\hat{r}}$,
can be derived directly
from the Einstein equations in the $\boldsymbol{e}_{\hat{r}},~ 
\boldsymbol{e}_{\hat{\theta}}, ~\boldsymbol{e}_{\hat{\phi}}$
basis.  The quasievolution equations are
\begin{equation}
D_0~K_R-D_{\hat{r}}~a_{\hat{r}}=S\line(1,0){1}K_R,
\end{equation}
\begin{equation}
D_0~K_T-D_{\hat{r}}~n_{\hat{r}}=S\line(1,0){1}K_T,
\end{equation}
\begin{equation}
D_0~n_{\hat{r}}-D_{\hat{r}}~K_T=S\line(1,0){1}n_{\hat{r}},
\end{equation}
with
\begin{multline}
S\line(1,0){1}K_R=a_{\hat{r}}^2-n_{\hat{r}}^2-K_R^2+K_T^2+
\frac{2~n_{\hat{r}}}{R},\\
\shoveleft{S\line(1,0){1}K_T=\frac{a_{\hat{r}}+n_{\hat{r}}}{R}+
K_R~K_T-K_T^2
-n_{\hat{r}}^2-a_{\hat{r}}~n_{\hat{r}},}\\
\shoveleft{S\line(1,0){1}n_{\hat{r}}=\frac{K_T-K_R}{R}~-~
a_{\hat{r}}~K_T~+~n_{\hat{r}}~(K_R-2~K_T).~~~~~~~~~~~~~~~~~~~~}\nonumber
\end{multline}
The momentum and energy quasiconstraint equations are, respectively,
\begin{equation}
\label{momquasicon}
D_{\hat{r}}~K_T=\frac{(K_R-K_T)~(1-R~n_{\hat{r}})}{R},
\end{equation}
\begin{equation}
\label{enquasicon}
D_{\hat{r}}~n_{\hat{r}}=\frac{3~n_{\hat{r}}^2}{2}-\frac{2~n_{\hat{r}}}{R}
-\frac{K_T}{2}~(2~K_R+K_T).
\end{equation}

The gauge quasievolution equations for $a_{\hat{r}}$ are not covariant 
under rotation, and must be derived from the Nester and Lorentz gauge 
conditions in a Cartesian basis (Eqs. (44) and (47) of \cite{BuBa03}).  
The results are converted to our spherical basis variables.
Both the Nester and Lorentz gauges give
quasievolution equations for $a_{\hat{r}}$ of the form 
\begin{equation}
D_0~a_{\hat{r}}-D_{\hat{r}}~K_R=S\line(1,0){1}a_{\hat{r}}.
\end{equation}
The source, $S\line(1,0){1}a_{\hat{r}}$ depends on the gauge. 
For the Nester gauge,
\begin{equation}
S\line(1,0){1}a_{\hat{r}}=\frac{2~(K_R-K_T)}{R},
\end{equation}
and for the Lorentz gauge,
\begin{equation}
\label{Lorentzsource}
S\line(1,0){1}a_{\hat{r}}=\frac{2~(K_R-K_T)}{R}
-2~(K_R~n_{\hat{r}}+K_T~a_{\hat{r}}).
\end{equation}
The Nester gauge quasiconstraint equation
is trivial because in spherical symmetry, the curl of a radial vector
is zero.

\section{True Evolution Equations and their Hyperbolic Structure}
\label{hyperbolichere}

In order to evolve $K_R,~K_T,~n_{\hat{r}},$ and $a_{\hat{r}}$ numerically,
the evolution equations must be expressed as partial derivatives along the 
coordinate directions $r$ and $t$.  To obtain these
true evolution equations, Eqs. (\ref{rhatdir}) and (\ref{transvdir}) are 
substituted into the quasievolution
equations in Sec. \ref{sectetradeqn}.  
Linear combinations of the quasievolution 
equations are taken to isolate the time derivative of each variable.
The results can be lumped together in the form
\begin{equation}
\label{trueevoln}
D_0~{\bf q}+\boldsymbol{C}^{\hat{r}}~B_R~\partial_r~{\bf q}={\bf S}.
\end{equation}
In this equation,
\begin{equation}
{\bf q} = \left ( \begin{array}{c}
K_R \nonumber \\
a_{\hat{r}} \nonumber \\
K_T \nonumber \\
n_{\hat{r}} \end{array} \right ),
~~~~~~~~~~\boldsymbol{C}^{\hat{r}}=-~\frac{1}{1-A_{\hat{r}}^2}\:\:
\left ( \begin{matrix}
        {A}_{\hat{r}}&1&0&0\\
        1&{A}_{\hat{r}}&0&0\\
        0&0&{A}_{\hat{r}}&1\\
        0&0&1&{A}_{\hat{r}},
\end{matrix} \right ),
\end{equation}
and
\begin{equation}
{\bf S}~~=~~\frac{1}{1-A_{\hat{r}}^2}\:\:
\left ( \begin{array}{c}
S\line(1,0){1}K_R+A_{\hat{r}}\:S\line(1,0){1}a_{\hat{r}} \nonumber \\
S\line(1,0){1}a_{\hat{r}}+A_{\hat{r}}\:S\line(1,0){1}K_R \nonumber \\
S\line(1,0){1}K_T+A_{\hat{r}}\:S\line(1,0){1}n_{\hat{r}} \nonumber \\
S\line(1,0){1}n_{\hat{r}}+A_{\hat{r}}\:S\line(1,0){1}K_T \end{array} \right ).
\end{equation}

The eigensystem of the characteristic matrix, $\boldsymbol{C}^{\hat{r}}$,
consists of four eigenmodes: two with eigenvalue $1/(1-A_{\hat{r}})$ and 
amplitudes $a_{\hat{r}}+K_R$ and  $n_{\hat{r}}+K_T$, and two with eigenvalue
$-1/(1+A_{\hat{r}})$ and amplitudes
$a_{\hat{r}}-K_R$ and $n_{\hat{r}}-K_T$.  Decomposing the $D_0$ operator 
into partial derivatives gives coordinates speeds
\begin{equation}
\label{s1}
s_1(r,t)=\frac{e^{-\lambda}~\alpha}{1+A_{\hat{r}}}-\beta^r,
\end{equation}
\begin{equation}
\label{s2}
s_2(r,t)=-\frac{e^{-\lambda}~\alpha}{1-A_{\hat{r}}}-\beta^r.
\end{equation}

The two eigenmodes involving $K_R$ and $a_{\hat{r}}$ , are ``longitudinal''
modes, since they are constructed from components of the extrinsic curvature
tensor and acceleration vector projected
along $\boldsymbol{e}_{\hat{r}}$.  The two
involving $K_T$ and $n_{\hat{r}}$,
are ``constraint'' modes, since $K_T$ and $n_{\hat{r}}$ are the variables 
which appear in the principal parts of the constraint equations.

\section{True Constraint Equations}
\label{trueconstraints}

To eliminate the time derivatives hidden in the quasiconstraint equations,
we use the true evolution equations.  It is convenient to take linear 
combinations of the result to get decoupled equations for 
$B_R~\partial_r~K_T$ and $B_R~\partial_r~n_{\hat{r}}$. 
We call these the true momentum and energy constraint equations,
\begin{multline}
\label{truemom}
B_R~\partial_r~K_T=(K_R-K_T)\left(\frac{1}{R}-n_{\hat{r}}\right)\\
-\frac{A_{\hat{r}}}{2}\left[\frac{2}{R}~(a_{\hat{r}}-n_{\hat{r}})
-3K_T^2-2a_{\hat{r}}n_{\hat{r}}+n_{\hat{r}}^2\right],
\end{multline}
\begin{multline}
\label{trueenergy}
B_R~\partial_r~n_{\hat{r}}=-~\frac{2~n_{\hat{r}}}{R}-K_R~K_T-
\frac{K_T^2}{2}+\frac{3~n_{\hat{r}}^2}{2}\\
\shoveleft{~~~~+A_{\hat{r}}~K_T~(a_{\hat{r}}+n_{\hat{r}}).~~~~~~~~~~~~~~~~~~}
\end{multline}
These equations are used in calculating initial conditions, 
boundary conditions, and as an accuracy check for the numerical evolution.

\section{Evolution and Constraint Equations for $\boldsymbol{B}_{\hat{r}}$,
$\boldsymbol{B}_{\hat{\theta}}$,
and $A_{\hat{r}}$}
\label{Bconstraints}

Evolution equations for the triad vector components are also required.  
Recall there are only two independent components in spherical symmetry, 
$B_R = e^{-\lambda}$ and 
$B_T = 1/R$.  Using $B_T$ rather than $R$ as
a variable in the numerics is motivated by a desire to maintain a
form of the equations similar to what is necessary in a three-dimensional 
calculation, but also leads to a significant improvement 
in accuracy of the results for the Lorentz gauge.  The
evolution equations for
$\boldsymbol{B}_{\hat{r}}$ and
$\boldsymbol{B}_{\hat{\theta}}$ are
\begin{multline}
\label{lambdaevolution}
(\partial_t - {\mathcal {L}}_{\boldsymbol{\beta}})\boldsymbol{B}_{\hat{r}}
=-~\alpha\:K_R\:\boldsymbol{B}_{\hat{r}}\\
\:\Rightarrow\:D_0~B_R=-~(K_R+\frac{\partial_r {\beta}^r}{\alpha})~B_R,
\end{multline}
\begin{multline}
\label{Revolution}
(\partial_t - {\mathcal {L}}_{\boldsymbol{\beta}})\boldsymbol{B}_{\hat{\theta}}
=-~\alpha\:K_T\:\boldsymbol{B}_{\hat{\theta}}\\
\:\Rightarrow\:D_0~B_T=-~K_T~B_T.~~~~~~~~~~~~~
\end{multline}
There is a constraint equation for $B_T$,
\begin{equation}
\label{Rconstraint}
B_R~\partial_r~B_T=-~(B_T-n_{\hat{r}}-A_{\hat{r}}~K_T)~B_T,
\end{equation}
which is equivalent to Eq. (\ref{Rprimeconstraint}) for the coordinate 
derivative of $R$.  This constraint is used to obtain $B_T$ 
(and $R$) in the initial conditions, and as an additional 
check on the accuracy of the numerical evolution.
The evolution equation for $A_{\hat{r}}$ follows from commuting $D_0$ with the
spatial directional derivatives,
\begin{equation}
\label{Arevoln}
D_0~A_{\hat{r}}=a_{\hat{r}}-K_R~A_{\hat{r}}
-B_R~\partial_r (\ln{\alpha}).
\end{equation}

\section{The Initial Value Problem (IVP)}

The initial value problem consists of finding values
for $A_{\hat{r}}$, $K_T$, $n_{\hat{r}}$, $B_T$, $B_R$,
$K_R$, and $a_{\hat{r}}$ with which to begin the
numerical evolution.  We take the congruence orthogonal
to the initial hypersurface, so $A_{\hat{r}}=0$ initially.
$K_T$, $n_{\hat{r}}$, and $R$ are obtained from Eqs. (\ref{truemom}), 
(\ref{trueenergy}), and (\ref{Rconstraint}) with  
$A_{\hat{r}}=0$.
The first integral of the constraint equations,
\begin{equation}
\label{firstintegral}
(1-R~n_{\hat{r}})^2-K_T^2=1-\frac{2M}{R},~~~
\end{equation}
where the constant of integration $M$ is the Schwarzschild mass,
makes numerical integration of Eq. (\ref{trueenergy}) for $n_{\hat{r}}$
unnecessary.  Using Eq. (\ref{firstintegral}) rather than 
Eq. (\ref{trueenergy}) to obtain $n_{\hat{r}}$ in the initial conditions 
has a substantial impact on the constraint errors early in the evolution, 
but after a few dynamical times leads to only a modest improvement in 
accuracy.

The free initial data are $K_R$, $a_{\hat{r}}$, and $B_R$ 
on the initial hypersurface.  The choice of $B_R$ is relatively
trivial, since it sets the initial relationship of coordinate
radius to proper radius and plays no role in the subsequent dynamics
of the congruence or the geometry.  With grid spacing uniform in
coordinate radius, the initial choice of $B_R$
can affect numerical accuracy simply because it 
affects the relative grid resolution in different parts of the 
domain.  Two particular prescriptions for 
$K_R$ and $a_{\hat{r}}$ are considered:
one which, along with appropriate choices
for the lapse and shift, leads analytically to a time independent 
solution of the evolution equations (Time Independent IVP), and
one which sets a uniform value for the trace of the extrinsic 
curvature on the initial hypersurface (Constant Mean Curvature IVP).

\subsection{Time Independent IVP}
\label{IVP}

The Schwarzschild spacetime has a time Killing vector field 
(timelike outside the horizon, $R > 2M$),
which means that coordinate systems can be found in which
the metric is time independent.  However, in the WEBB tetrad
formulation the variables will be time independent only 
if the tetrad congruence is stationary as well as the (coordinate) 
metric.  Reproducing an analytically time-independent solution is
the simplest and most basic test of a numerical code.  Note that 
the standard Schwarzschild slicing giving rise to a static metric is 
not satisfactory for our purposes, since the normal tetrad congruence 
would be singular, with infinite acceleration, on the horizon.

In a time-independent evolution, a
congruence initially orthogonal to a constant-time hypersurface will be 
orthogonal at all times, so $A_{\hat{r}}$ will be zero at all times.  This 
is consistent with Eq. (\ref{Arevoln}) if the lapse is given by
\begin{equation}
\label{lapsecondhere}
B_R~\partial_r~(\ln{\alpha})=~a_{\hat{r}}.
\end{equation}
Also, the coordinate time derivative of the geometrically defined 
curvature radius $R$ must be zero, 
a condition which, by Eqs. (\ref{Revolution}) 
and (\ref{Rconstraint}), requires that the shift satisfy
\begin{equation}
\label{shiftcond}
\beta^r=- \, \frac{\alpha~B_R~R~K_T}{1-R~n_{\hat{r}}}.
\end{equation}
With $\partial_t~B_R=0$, $B_R~\partial_r~(\beta^r/B_R)=\,-\alpha~K_R$.

Setting the partial time derivatives of
$K_R$ and $a_{\hat{r}}$ to zero
in their true evolution equations gives two simultaneous 
equations for the proper radial derivatives of $K_R$ and $a_{\hat{r}}$.  
The equation for $K_R$ is 
\begin{multline}
\label{diffKR}
B_R~\partial_r~K_R=\,-\,(1-R~n_{\hat{r}})\\
\times\frac{R~K_T~S\line(1,0){1}K_R+(1-R~n_{\hat{r}})
~S\line(1,0){1}a_{\hat{r}}} 
{(1-R~n_{\hat{r}})^2-(R~K_T)^2}\,.
\end{multline}
Note that $S\line(1,0){1}a_{\hat{r}}$ depends on the choice of tetrad gauge, 
{\it{ie.}} Nester versus Lorentz.

Eq. (\ref{diffKR}) is singular when $(1-R~n_{\hat{r}})^2-(R~K_T)^2=0$,
or $R=2M$ (see Eq. (\ref{firstintegral})). 
A solution regular on the horizon is obtained by requiring that the
numerator of Eq. (\ref{diffKR}) vanish at $R=2M$, which implies a 
relation between the values of $K_R$ and $R~K_T$ on the horizon 
(which we call $K_{RH}$ and $U_0$, respectively).
For the Nester gauge, this relation is
\begin{equation}
2~M~K_{RH}=\frac{1+4~{U_0}^2-8~\left|U_0\right|^3}
{-4~U_0-8~U_0~\left|U_0\right|}~,
\end{equation}
and for the Lorentz gauge, it is
\begin{equation}
2~M~K_{RH}=\frac{-1+8~\left|U_0\right|^3}{4~U_0}~.
\end{equation}
$U_0$ is a free parameter, which must be 
negative for a black hole horizon.  
L'H{\^o}pital's rule can be used to find the 
starting value for $B_R~\partial_r~K_R$ on the horizon.

While one can integrate the equation for $B_R~\partial_r~a_{\hat{r}}$ obtained 
along with Eq. (\ref{diffKR}), it is simpler to use the algebraic 
expression
\begin{equation}
\label{aralg}
a_{\hat{r}}=\frac{M/R^2+R~K_T~K_R}{(1-R~n_{\hat{r}})}.
\end{equation}
Eq. (\ref{aralg}) is obtained by requiring that 
$\partial_t(R~K_T)=0$, eliminating radial derivatives of 
$R~K_T$ and $n_{\hat{r}}$ using the constraint equations,
eliminating the shift using Eq. (\ref{shiftcond}), and finally, 
simplifying with the help of Eq. (\ref{firstintegral}).
With $a_{\hat{r}}$ given by Eq. (\ref{aralg}),
we can determine the lapse everywhere
using Eq. (\ref{lapsecondhere}).

In summary, for a particular choice of $U_0$, 
one can obtain $K_R$, $a_{\hat{r}}$,
$\alpha$, and $\beta^r$ everywhere so that the evolution
of all the variables is time independent.  The constant-time 
hypersurface generally does become 
singular at some point inside the horizon, so the 
initial hypersurface must be terminated at an excision boundary inside 
the horizon before $a_{\hat{r}}$ and/or $K_R$ become too large.
At large $R$, the generic behavior is that $K_T$ and $K_R$ approach the 
same constant value.  This constant is positive if $U_0$ is 
greater (less negative) than a certain critical value, which is exactly
$-0.25$ for the Nester gauge and about $-0.29$ for the Lorentz gauge.  A
positive $K_T$ at large $R$ is highly desirable in dealing with outer 
boundary conditions, because then the shift at that outer boundary is 
negative and modes propagating along the congruence propagate out of,
rather than into, the grid.  Also, we find that
expansion of the congruence along the 
radial direction ($K_R > 0$) everywhere is generally helpful 
in reducing growth 
rates of any unstable constraint-violating modes.

\subsection{Constant Mean Curvature (CMC) Slice}
\label{cmc}

An attractive slicing condition from the point of view of the conformal 
approach to the initial value problem 
\cite{BowenYork80,York99,PfeifferYork2003} is to 
impose a uniform value for the trace of the extrinsic curvature of the
initial hypersurface.  This ``Constant Mean Curvature'' slicing, with
$(\text{trace}~K=K_R+2~K_T=K_0)$, allows testing of cases where the evolution of the
hypersurfaces and the congruence is time dependent.  In order to make the 
evolution of the congruence as ``quiet'' as possible, we choose the initial
acceleration to satisfy the stationarity condition of Eq. (\ref{aralg}).  
Then, with $K_R=K_0-2K_T$, the momentum and energy constraints can be
integrated as differential equations for $K_T$ and $n_{\hat{r}}$, starting 
at the horizon, as described in the previous section.  
$U_0$ (the value of $R~K_T$ 
on the horizon) is a free parameter, along with the choice of $K_0$.  The
value of $n_{\hat{r}}$ on the horizon is fixed by Eq. (\ref{firstintegral}).

Together with a shift given by Eq. (\ref{shiftcond}) and a lapse given 
by Eq. (\ref{lapsecondhere}), this implementation of the CMC initial 
condition means that the time derivatives of 
$K_T$ and $n_{\hat{r}}$ vanish on the initial hypersurface, though they 
will not stay zero.  A critical test 
of the Nester and Lorentz tetrad gauge conditions is whether the 
hypersurfaces and congruence will subsequently evolve toward or 
away from a time-independent solution.  Of course, the answer to this question
also depends on how the 
coordinates evolve, as described in the next section.

\section{Resetting Coordinate Conditions}
\label{resetting}

Our system of equations is symmetrizable hyperbolic for any fixed choice
of lapse and shift.  Accordingly, during
a given time step, the lapse and shift
are held fixed at the values they have
at the start of the time step.
While it is essential to accuracy and stability
that the system be hyperbolic during the 
time steps \cite{BaBu02}, it is not critical
that the overall evolution be hyperbolic,
only well-posed \cite{Reula98}.
In other words, the lapse and the shift can be reset at fixed
time intervals according to conditions which
do not necessarily preserve the
hyperbolicity of the system.
This gives a wide range of choices
for resetting conditions.  Keeping a fixed lapse and shift for many 
dynamical times is likely to give rise to coordinate singularities.

The lapse determines the evolution of the constant-time hypersurfaces.
The hyperbolic system breaks down if $A_{\hat{r}}=1$, which signifies
that the constant-time hypersurface has become null.  We reset the 
lapse to keep $A_{\hat{r}}$ small, which is accomplished by making
the reset lapse satisfy Eq. (\ref{lapsecondhere}).  Then the time
derivative of $A_{\hat{r}}$ will be small by Eq. (\ref{Arevoln}) as
long as $A_{\hat{r}}$ is small.  Since $A_{\hat{r}}=0$ initially, it
does in fact stay very small if the lapse is reset at small time
intervals.  The constant of integration in solving Eq. 
(\ref{lapsecondhere}) is conveniently chosen to keep the lapse 
constant at the outer edge of the grid, but makes no practical 
difference, since it is just a uniform rescaling of the time 
coordinate.  Choosing the time step according to a Courant condition 
automatically compensates for any such rescaling.

The shift controls the evolution of the spatial coordinates.  We work 
with a grid at fixed values of the radial coordinate.  The evolution 
of the radial coordinate should be managed to keep the grid from being 
sucked up by the black hole, while keeping the 
inner edge just inside the event 
horizon, without any excessive stretching or compression of the grid
relative to the physical curvature radius $R$.  If 
made possible by the evolution of the congruence and the constant-time 
hypersurfaces, we want our variables to approach a stationary final 
state.  One option would be to choose the shift to make the partial 
time derivative of the curvature radius $R$ zero everywhere at
each resetting.  From Eqs. (\ref{Revolution}) and (\ref{Rconstraint}) 
this condition is
\begin{equation}
\label{constRshift}
\beta^r=-\frac{\alpha~B_R~R~K_T}{1-R~n_{\hat{r}}-
~A_{\hat{r}}~R~K_T}.
\end{equation}
However, it seems more desirable to use a condition 
which is not so tied to the special circumstances of spherical 
symmetry.  

The minimal strain condition, as introduced by Smarr and York \cite{SY78}
for $3+1$ formulations, minimizes an integral of the square of
the time derivative of the spatial metric over the hypersurface,
\begin{equation}
\label{action}
I_1=\int \left[\frac{\partial h_{ij}}{\partial t}~
\frac{\partial h_{kl}}{\partial t}~h^{ik}~h^{jl}\right]~
\sqrt{h}~d^3x
\end{equation}
with respect to variations in the shift.
In $I_1$, $h_{ij}$ is the spatial metric and $h$ is the determinant 
of this metric.  The metric time derivatives depend on the shift 
through
\begin{equation}
\frac{\partial h_{ij}}{\partial t}=-2~\alpha~K_{ij}
+{\mathcal {L}}_{\boldsymbol{\beta}}~h_{ij},
\end{equation}
where $K_{ij}$ is the extrinsic curvature of the hypersurface.

In the context of the WEBB equations, the vectors $\boldsymbol{B}_a$ carry the
information corresponding to the spatial metric in a 3+1 formalism.  
These vectors are not unit vectors with respect to the true spatial 
metric if the tetrad 
congruence is not orthogonal to the constant-time hypersurface.  
As projections of the spatial triad vectors into the constant-time 
hypersurface, they can be 
thought of as representing unit spatial displacements in the local 3-space 
orthogonal to the tetrad congruence, which are Lorentz-contracted when
measured in the hypersurface frame.  We find it convenient to pose a minimal
deformation condition on the components $B_a^{~k}$, 
which in general is {\it{not}} the same as the Smarr-York condition.

Inverting the $B_a^{~k}$ matrix gives the matrix $B^a_{~k}$ 
of components of one-forms dual to the $\boldsymbol{B}_a$ vectors.  
A spatial metric with respect to which the $\boldsymbol{B}_a$ 
are unit vectors is $h_{ij}=B^a_{~i}~B^a_{~j}$.  We use this
metric, rather than the true spatial metric, in our minimal deformation
condition, minimizing with respect to variations in the shift vector the action
\begin{equation}
\label{EWaction}
I_2=\int h_{ij}~\frac{\partial \, B_b^{~i}}{\partial t}~\frac{\partial 
\, B_b^{~j}}{\partial t}~\sqrt{h}~d^3x,
\end{equation}
where $h$ is the determinant of $h_{ij}$.
In the current application, with $A_{\hat{r}}$ 
kept very small by resetting the 
lapse and a diagonal $B_a^{~k}$, $I_2$ is very nearly equivalent to $I_1$.

The dependence on the shift is through
\begin{equation}
\label{ms1}
\frac{\partial_t B_R}{B_R}=\beta^r~\frac{\partial_r B_R}{B_R}-\alpha~K_R
-\partial_r~\beta^r \equiv y
\end{equation}
and
\begin{equation}
\label{RdotR}
\frac{\partial_t B_T}{B_T}=\beta^r~\frac{\partial_r B_T}{B_T}-\alpha~K_T.
\end{equation}
with
\begin{equation}
I_2=\int dr~4\pi~\frac{R^2}{B_R}~\left[\left(\frac{\partial_t B_R}
{B_R}\right)^2+2\left(\frac{\partial_t B_T}{B_T}\right)^2\right].
\end{equation}
The Euler-Lagrange equation is
\begin{equation}
\label{ms2}
\partial_r~y=2 \frac{\partial_r B_T}{B_T}
~\left(y+\beta^r~\frac{\partial_r B_T}{B_T}-~\alpha~K_T\right),
\end{equation}
which, along with Eq. (\ref{ms1}) giving $\partial_r\beta^r$ in terms of $y$,
can be integrated to find $\beta^r$ and $\partial_r\beta^r$ everywhere.  
Note that $\partial_r B_T$ can be evaluated from the constraint equation 
(\ref{Rconstraint}).  Strict minimization requires $y=0$ at 
each boundary, but in order to keep the inner edge of the grid just inside the
horizon and the outer edge at a fixed $R$ we instead 
impose Eq. (\ref{constRshift})
as the shift boundary condition at both edges of the grid.
After each recalculation the values of $\beta^r$ 
and $\partial_r\beta^r$ at each grid point are stored and kept fixed until 
the next resetting of the lapse and shift.

\section{Boundary Conditions}
\label{BC}

We now discuss boundary conditions for the evolution equations.
The inner boundary of the grid is an excision boundary, kept at 
a roughly constant $R < 2M$, inside the event horizon, by the shift 
condition of the last section.  All characteristics of the evolution 
equations propagating at less than or equal to light speed (which is 
all of them in the WEBB framework) are then outgoing relative to the 
grid at the inner boundary, since inside the horizon the entire future
light cone is toward decreasing $R$.  All information needed to update the 
variables at the inner boundary can be obtained from the boundary 
values or upwind differencing.

The outer boundary is a different story.  We do not try to extend the
outer boundary all the way to infinite $R$, though this in principle is
possible.  Since our asymptotically hyperbolic 
hypersurfaces are asymptotically null, the variables should be 
regular functions of $B_T=1/R$, and the radial 
coordinate could be scaled to be linear in $B_T$ approaching null infinity.
We terminate our initial grid at a value of $R$ the order of $20~M$, 
where variables such as $n_{\hat{r}}$, $K_T$, and $K_R$ have 
reached nearly constant values, and keep the outer
boundary at approximately the same $R$ with our shift condition.
This allows the grid to be
roughly evenly spaced in proper radius, with adequate resolution near the
horizon and a reasonable total number of grid points.  As long as $K_T$ is
positive near the outer boundary, the shift is negative there 
(see Eq. (\ref{constRshift})), and the 
modes propagating along the hypersurface normal are outward relative to
the boundary.  This is critically important, since there is no good way
of specifying incoming boundary conditions for these modes.  Still, the
two modes with speeds $s_2$ are incoming at the boundary and do require 
boundary conditions.  The amplitudes of these
modes are $(a_{\hat{r}}+K_R)$ and $(n_{\hat{r}}+K_T)$.
Hence, we refer to them as the incoming ``longitudinal''
and ``constraint'' modes, respectively.
The ``constraint'' modes consist
of the variables whose radial derivatives appear in the principal parts 
of the energy and
momentum constraint equations, and have the potential of propagating
constraint errors in from the boundary (see \cite{BaBu02} for
a fuller explanation).  The constraints cannot be satisfied by just
setting the incoming constraint mode amplitude to zero.  
A number of different constraint-preserving boundary conditions have
been proposed \cite{Stewart98,FN99,IR02,BaBu02,SW03,CPRST03,SSW02,CLT02,CS03,
frittelligomez04,LSKPST04,KLSBP05,ST05}.  What we do is, first, 
iteratively correct the values 
of the constraint variables at the boundary physical points so that 
the energy and momentum constraints evaluated at half a step inside 
the grid are satisfied.  Then all variables are extrapolated 
to the ghost point just outside the boundary, and the energy and 
momentum constraints are again applied iteratively halfway between the 
boundary physical point and the ghost point to adjust the
extrapolated values of the constraint variables.

For stationary initial conditions,
the longitudinal variables, $K_R$ and $a_{\hat{r}}$,
are kept equal to their initial values at
the ghost point, to give boundary conditions for
the incoming longitudinal mode,
$K_R+a_{\hat{r}}$.  
For CMC initial conditions, the ghost point longitudinal variables 
can either be extrapolated quadratically or cubically to the ghost point,
which could be dangerous for stability, or again
maintained at their initial values.  The incoming longitudinal mode 
amplitude is a purely gauge quantity, so the only real concern is not to 
do something which leads to a gauge instability at the boundary.

\section{Numerical Methods}

The numerical methods used
in the codes are designed
to be second order accurate in
both space and time.  While we have experimented with various methods,
the most accurate results, and the ones presented in this paper, were
obtained with a second order Strang split method.
During each full time step, the source terms alone are integrated for a
half time step at each grid point using second-order Runge-Kutta, 
followed by a full wave propagation time step with the updated variables 
solving the equations with source terms omitted, and then another 
half time step of source term integration.  In the wave propagation 
time step the equations are reshuffled into decoupled advection equations 
for the eigenmode amplitudes, of the form
\begin{equation}
\partial_t y + v~\partial_r y = 0,
\end{equation}
where $v$ is the wave speed.
Using characteristic tracing, 
the solution of the advection equation for the updated $y$ at the $i^{th}$ 
grid point is just the initial value of $y$ at the position 
$r_i - \bar{v}~\Delta t$, where $\bar{v}$ is 
the average advection velocity along
the characteristic.  This average speed is approximated by advancing
$v$ to halfway through the time step by a first-order accurate Euler
method at all grid points before doing the wave propagation (only
necessary for the modes whose wave speed depends on $B_R$), and then
interpolating $1/v$ to the midpoint in $r$ along the characteristic during the
wave propagation, so at the $i^{th}$ grid point
\begin{equation}
\bar{v} = v_i /(1 \mp \frac{\Delta t}{2~\Delta x}(v_{i \mp 1} - v_{i})),
\end{equation}
with upper and lower signs for $(v_{i}>0)$ and $(v_{i}<0)$, respectively.
Standard second-order methods interpolate $y$
to the characteristic location using an upwind first-order difference and a 
second-order difference either centered at $i$ (Lax-Wendroff) or centered 
at the upwind grid point (Beam-Warming).  Except at the boundaries, we 
use a linear combination
of the two second-order differences which does the interpolation to
third-order accuracy, weighting the centered second-order 
difference by $(2-|v|)/3$ and the upwind second-order difference
by $(1+|v|)/3$.  
While the overall method is not 
third-order accurate, this does significantly reduce the errors compared
with pure Lax-Wendroff.
At the boundaries either pure Lax-Wendroff or pure Beam-Warming is used
as necessary to minimize use of ghost cells, Beam-Warming for outgoing
modes and Lax-Wendroff for incoming modes.
Note that the upwind point changes as $v$ changes sign, which means a
small but sudden change in the interpolated upwind second-order difference 
and its contribution to the updated $y$.  The effect of this is noticeable
in the constraint error plots we present in Sec. \ref{results}, but 
has no impact on stability.

The integration of the spatial ordinary differential equations
during the initial condition and boundary condition subroutines
is implemented with a simple second-order Predictor-Corrector
scheme \cite{Press86}, iterated
until the differences between
the predicted and corrected values
are all near machine precision.

\section{Results}
\label{results}

All our numerical results are obtained on a uniform grid ranging 
from $-0.16 \le r \le 9.84$, where $r$ is the coordinate radius,
with $r=0$ at the initial location of the event horizon.
The initial $B_R=1$, so that the coordinate
radius starts out equaling the proper radius.
The natural logarithm of the lapse ($\ln \alpha$), the shift,
($\beta^r$), and their derivatives are reset 
as described in Sec. \ref{resetting}.
For convergence studies, these quantities are reset at a constant 
coordinate time interval as the grid resolution is increased and
the time step is decreased in accordance with the Courant condition.
Note that all values plotted are in units with $2~M=1$.

\subsection{Time}

When presenting results,
it is desirable to relate the coordinate time ($t$) of the
numerical evolution to a physically meaningful time.  The 
natural choice for a physical time is the proper time of an
observer at infinite distance from the black hole, i.e., the 
Schwarzschild time coordinate $t_S$.  Our spacelike hypersurfaces 
are quite different from the constant-time surfaces in the 
Schwarzschild metric, but we can still relate changes in our
time coordinate to changes in Schwarzschild time.  

The derivation is outlined assuming $A_{\hat{r}}=0$,
since $A_{\hat{r}}$ is kept close to zero with our
lapse resetting condition.  The proper time for a displacement 
perpendicular to the two-spheres as obtained 
from the Schwarzschild metric is
\begin{equation}
d\tau=\sqrt{\left(1-\frac{2~M}{R}\right)
~dt_S^2-\frac{1}{1-\frac{2~M}{R}}~dR^2}.
\end{equation}
from which, with $dt$ the change in our coordinate time,
\begin{equation}
\left(\frac{dt_S}{dt}\right)^2
=\frac{1}{1-\frac{2~M}{R}}\left[\left(\frac{d\tau}{dt}\right)^2
+\frac{1}{1-\frac{2~M}{R}}~\left(\frac{dR}{dt}\right)^2\right].
\end{equation}
For a displacement at constant coordinate radius $r$ the metric of 
Eq.(\ref{spacetimemetric}) gives 
$(d\tau/dt)^2=(\alpha^2-({\beta^r~e^\lambda})^2)$.
At both edges of the grid our boundary condition 
on the minimal deformation shift  (Eq. \ref{constRshift}) 
gives $\partial_t~R=0$, and simplifying with the 
help of Eq. (\ref{firstintegral}) we arrive at 
\begin{equation}
dt_S=\frac{\alpha}{1-R~n_{\hat{r}}}~dt ~~~
\end{equation}
for the relation between between Schwarzschild time and coordinate time there.
We can normalize the lapse
at each resetting so that $\alpha=(1-R~n_{\hat{r}})$
at the outermost grid point, so that the
change in Schwarzschild time equals the change in coordinate
time at the outer edge of the grid to reasonable accuracy.
For stationary initial conditions, this also makes $dt_S=dt$ hold at all 
interior grid points.

\subsection{Stationary IVP}
\label{stationaryIVP}

Testing the code using a stationary solution as the
initial condition is the first step toward determining
the viability of the code.  Ideally, as the
numerical evolution proceeds, 
all the variables stay constant.  This does not occur
in practice, however, because numerical errors act as perturbations
to the stationary solution.  Even with an accurate and stable numerical
method, these perturbations may excite analytic unstable modes of the
evolution equations, some of which may be constraint-violating and some 
of which may be purely gauge instabilities.

\begin{figure*}[h]
\resizebox{2in}{2.1in}{\includegraphics{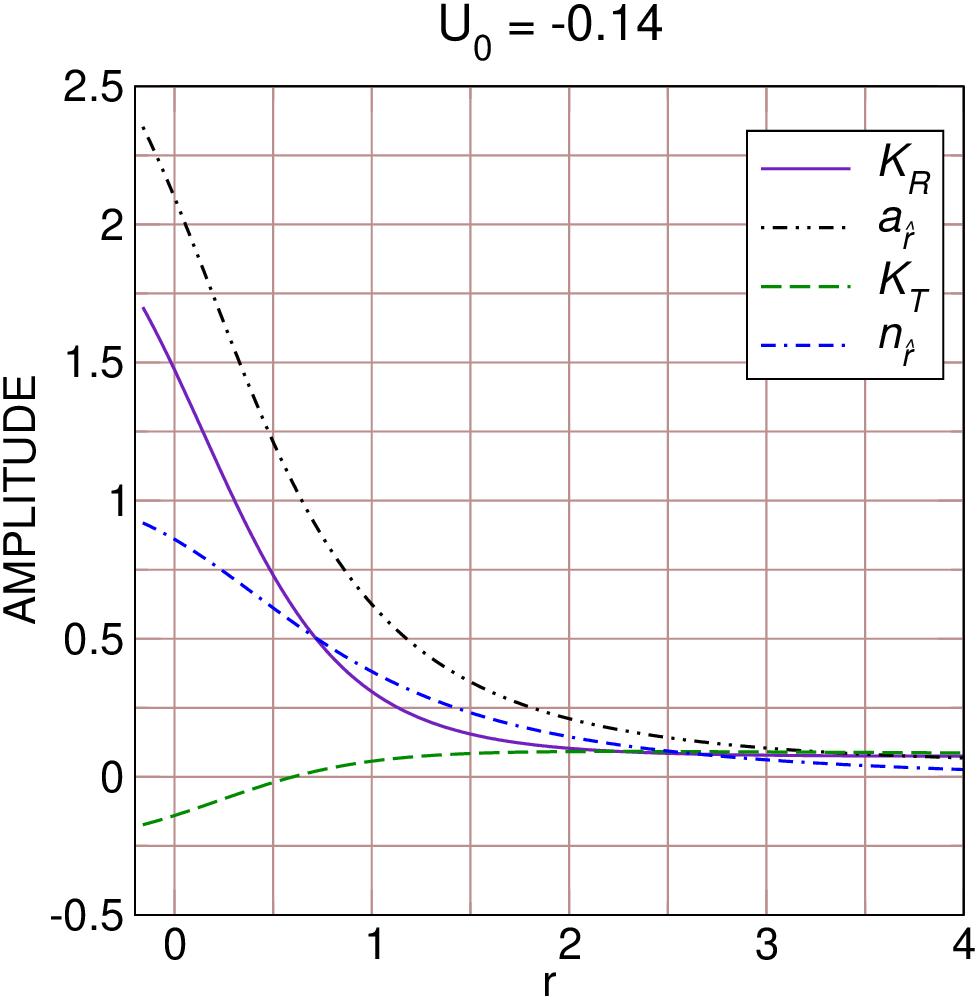}}
\hspace{.2in}
\resizebox{1.9in}{2.1in}{\includegraphics{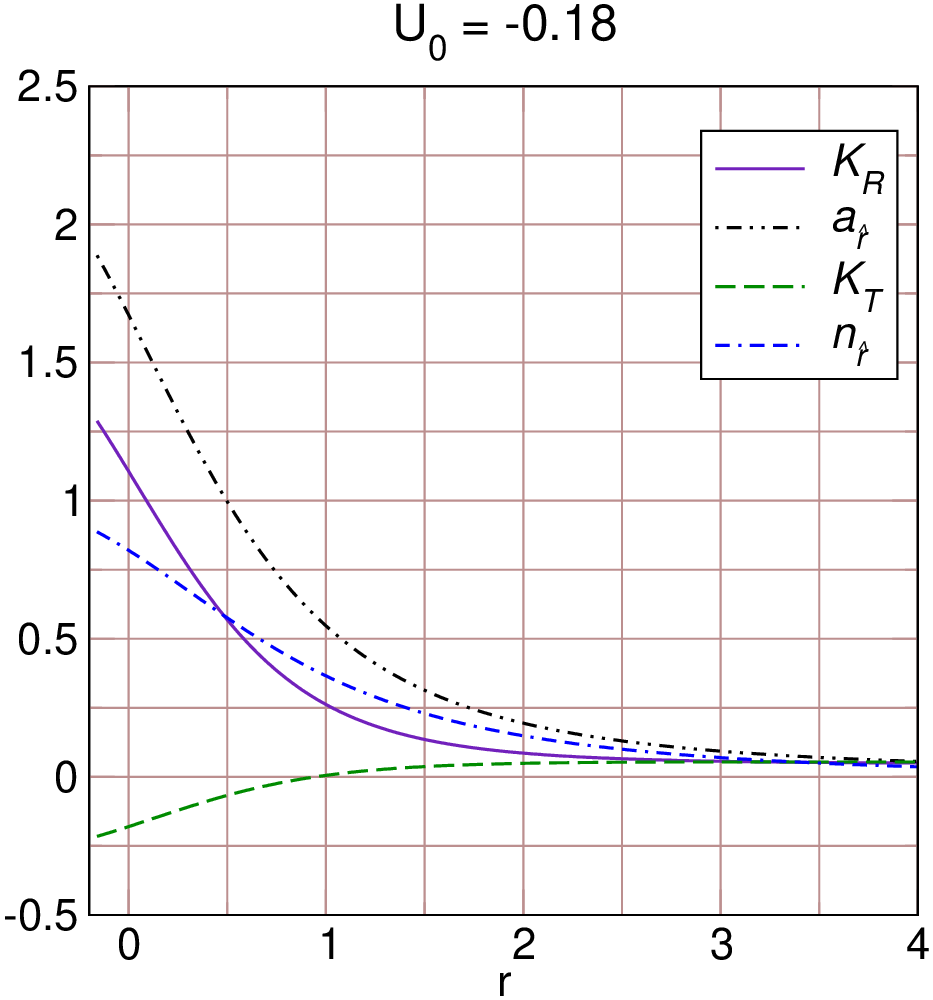}}
\hspace{.2in}
\resizebox{1.9in}{2.1in}{\includegraphics{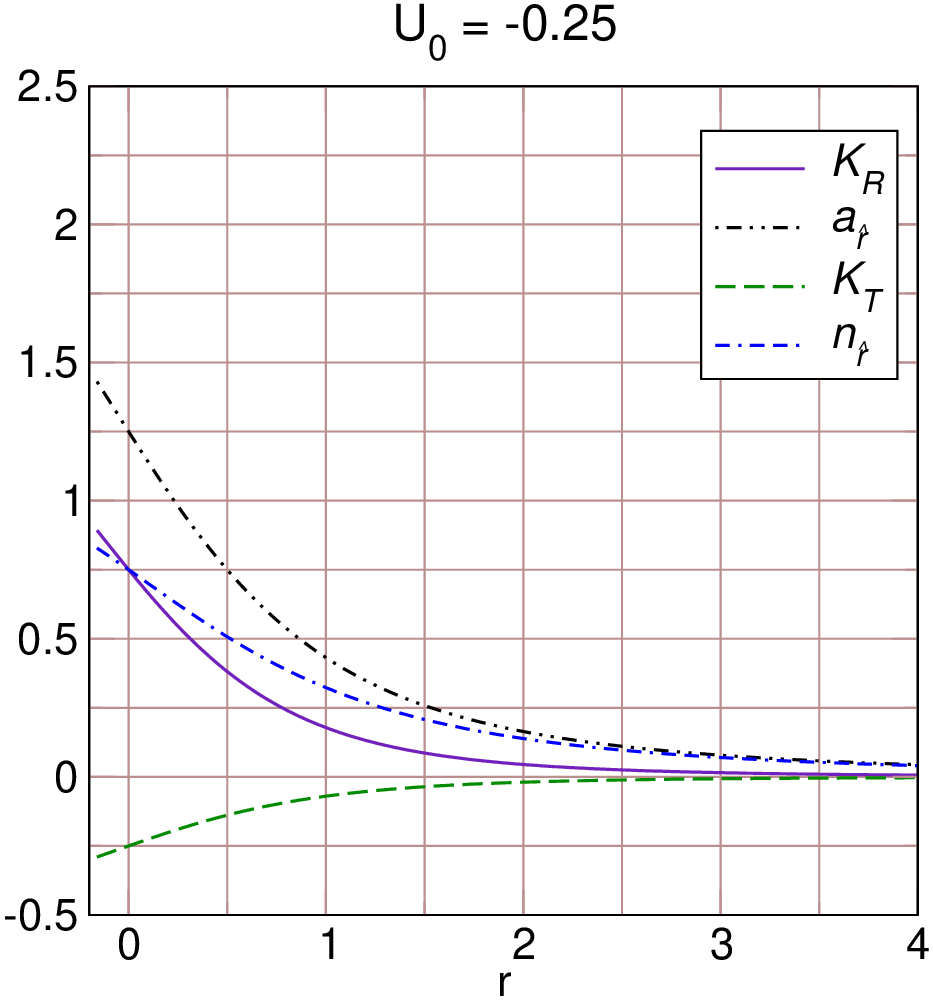}}
\caption{Stationary Schwarzschild solution, Nester gauge.  Shown is
the region $-0.16<r<4.0$.  The outer boundary is located at $r=9.84$.}
\vspace{2cm}
\label{stat_soln_Nester}
\end{figure*}

\begin{figure*}[h]
\resizebox{2in}{2.1in}{\includegraphics{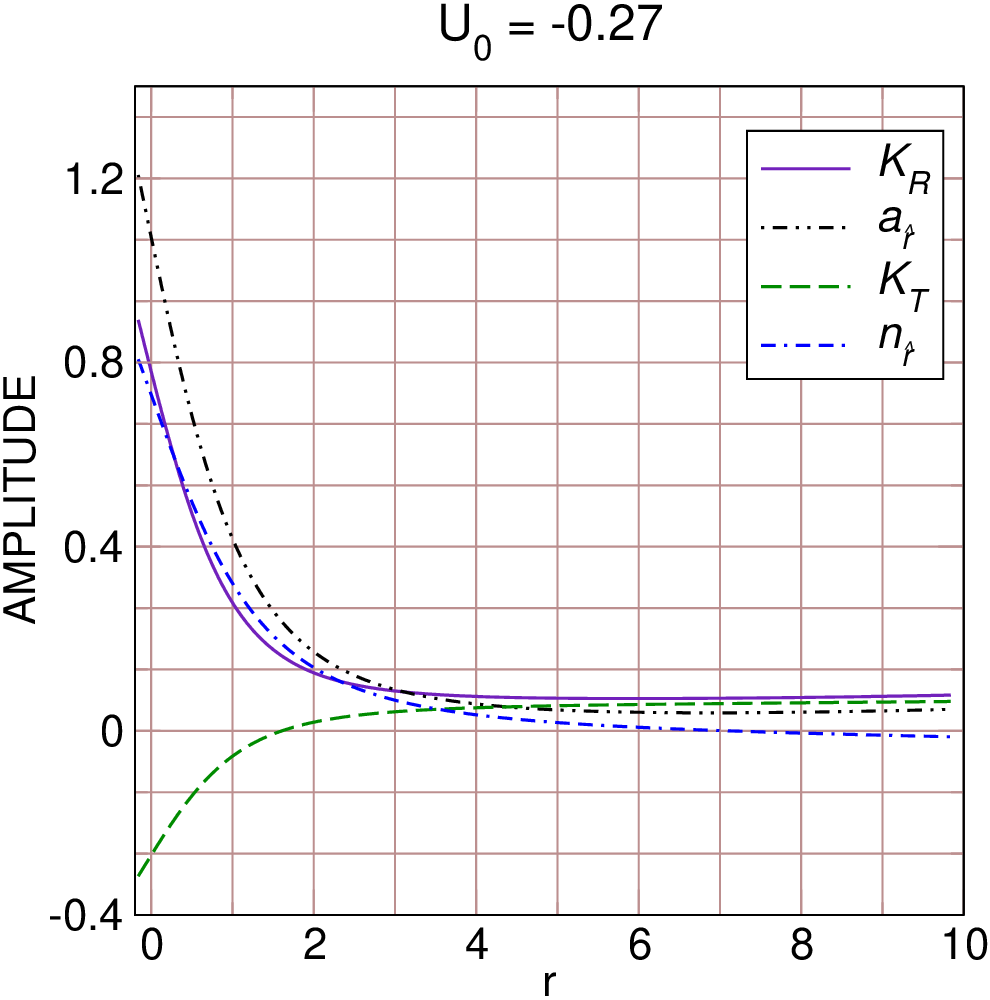}}
\hspace{.2in}
\resizebox{1.9in}{2.1in}{\includegraphics{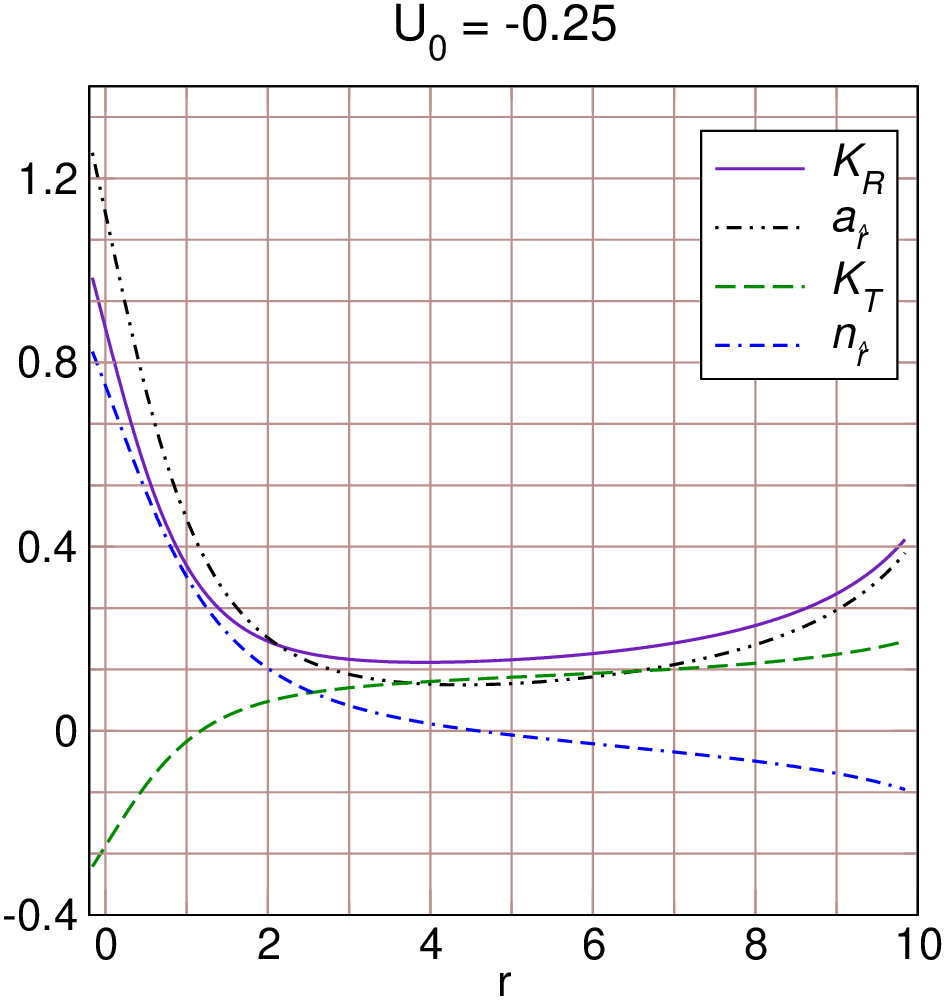}}
\caption{Stationary Schwarzschild solution, Lorentz gauge.}
\vspace{2cm}
\label{stat_soln_Lorentz}
\end{figure*}

Figs. \ref{stat_soln_Nester} and \ref{stat_soln_Lorentz} show 
solutions of the stationary initial value problem for the Nester and 
Lorentz gauges, respectively.  The solutions are one-parameter families 
characterized by $U_0$, the value of $R~K_T$ on the horizon.  We only 
consider cases where the radial component of the extrinsic curvature $K_R$ 
is positive (expansion of the normals radially) at large $r$, since only 
then do modes propagating along the hypersurface normal propagate off of, 
rather than onto, the grid at the outer boundary.  

For the Nester gauge, only the inner portion of the spatial domain is shown, 
from just inside the horizon at $r=-0.16$ to $r=4.0$, since the asymptotic 
behavior of most of the variables is already apparent by $r=4.0$.  The 
outer edge of the grid is always at $r=9.84$.  At the critical value 
$U_0=-0.25$ all of the variables plotted approach zero as $R \rightarrow 
\infty $, and the constant-time hypersurface asymptotically approaches a 
Schwarzschild constant-time hypersurface.  For $U_0$ less negative than 
the critical value, both $K_R$ and $K_T$ approach the same limiting 
(positive) value, with $\text{trace}~K \rightarrow 0.15$ for $U_0=-0.18$ and
$\text{trace}~K \rightarrow 0.24$ for $U_0=-0.14$.  The gradients in $K_R$ 
and $a_{\hat{r}}$ become steeper in the vicinity of the event horizon 
at $r=0$ as $U_0$ becomes less negative.  Note that the first-integral 
constraint of Eq. (\ref{firstintegral}) requires a change of sign of
$n_{\hat{r}}$ from positive to negative going outward when $K_T$ approaches 
a non-zero value at large $R$.  This occurs outside the range of these 
plots, but not too far outside for $U_0=-0.14$.

For the Lorentz gauge, we show the entire range of the grid, from just 
inside the horizon at $r=-0.16$ to the outer edge at $r=9.84$.  The 
critical value of $U_0$ at which the asymptotic values of $K_T$ and $K_R$ 
change sign is about $-0.29$.  The asymptotic behavior at large R when 
$U_0$ is less negative than the critical value is not well-behaved.  Both 
$K_R$ and $K_T$ get steadily larger, eventually in a runaway indicating 
that the stationary condition is forcing the hypersurface to become null.  
The beginning of this runaway is apparent in the plot for $U_0=-0.25$, but 
is well outside the edge of the grid for $U_0=-0.27$.  This indicates a 
pathology in the Lorentz gauge which does not 
allow an approach to a stationary 
solution over the entire spacetime, but may not cause problems when 
solutions are evolved over a limited domain.  The pathology is caused by 
the extra terms in the Lorentz gauge source for the $a_{\hat{r}}$ 
evolution equation (Eq. \ref{Lorentzsource}).  Once $n_{\hat{r}}$ becomes 
negative the $-2~K_R~n_{\hat{r}}$ term becomes positive and exerts positive 
feedback on $K_R$ through Eq. (\ref{diffKR}).

\begin{figure*}[h]
\resizebox{2in}{2.1in}{\includegraphics
{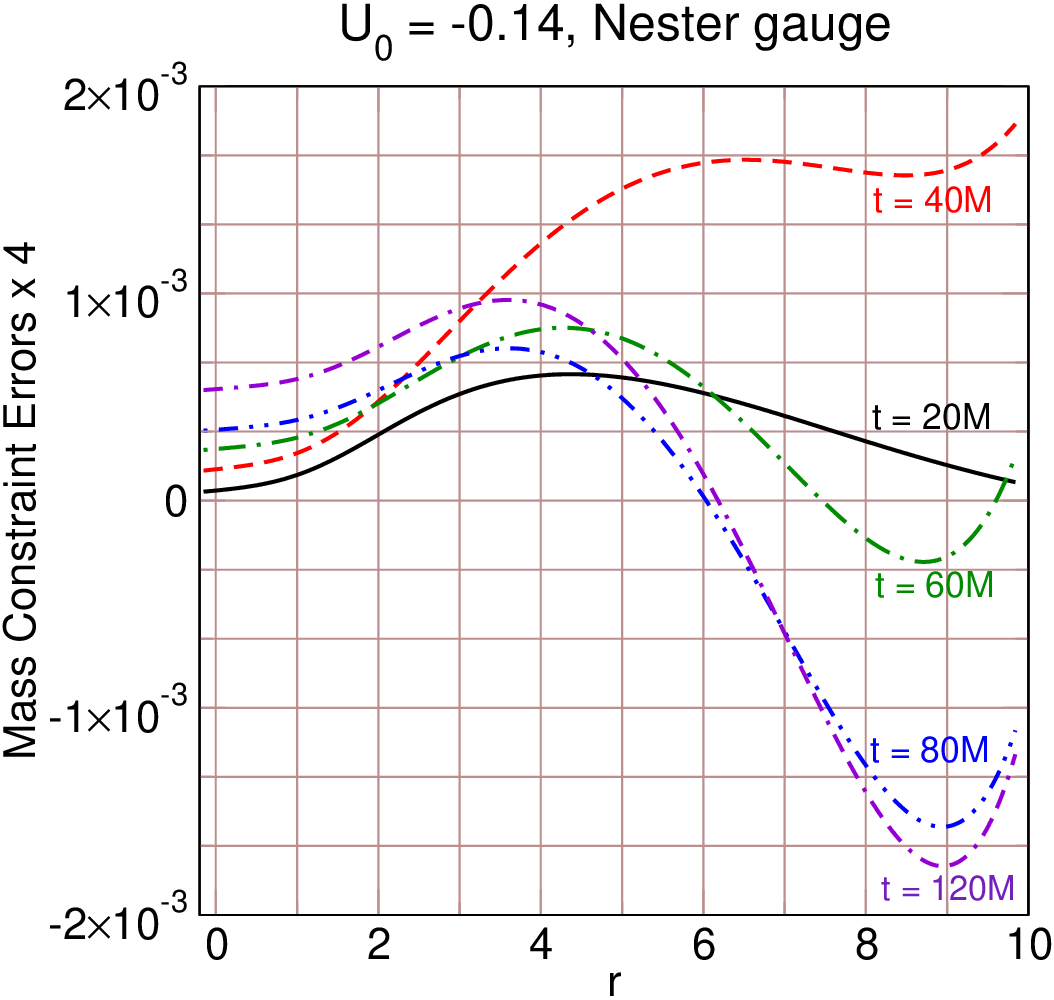}}
\hspace{.2in}
\resizebox{2in}{2.1in}{\includegraphics
{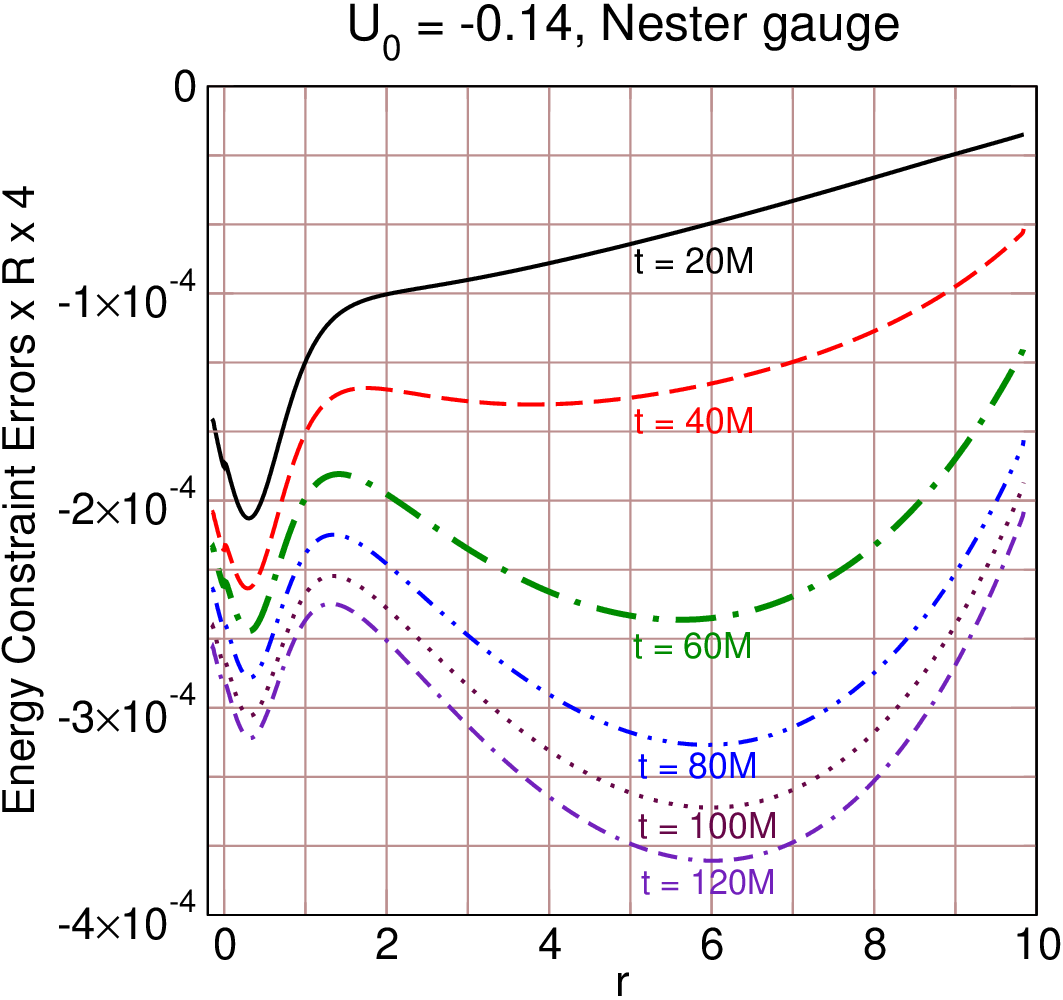}}
\hspace{.2in}
\resizebox{2in}{2.1in}{\includegraphics
{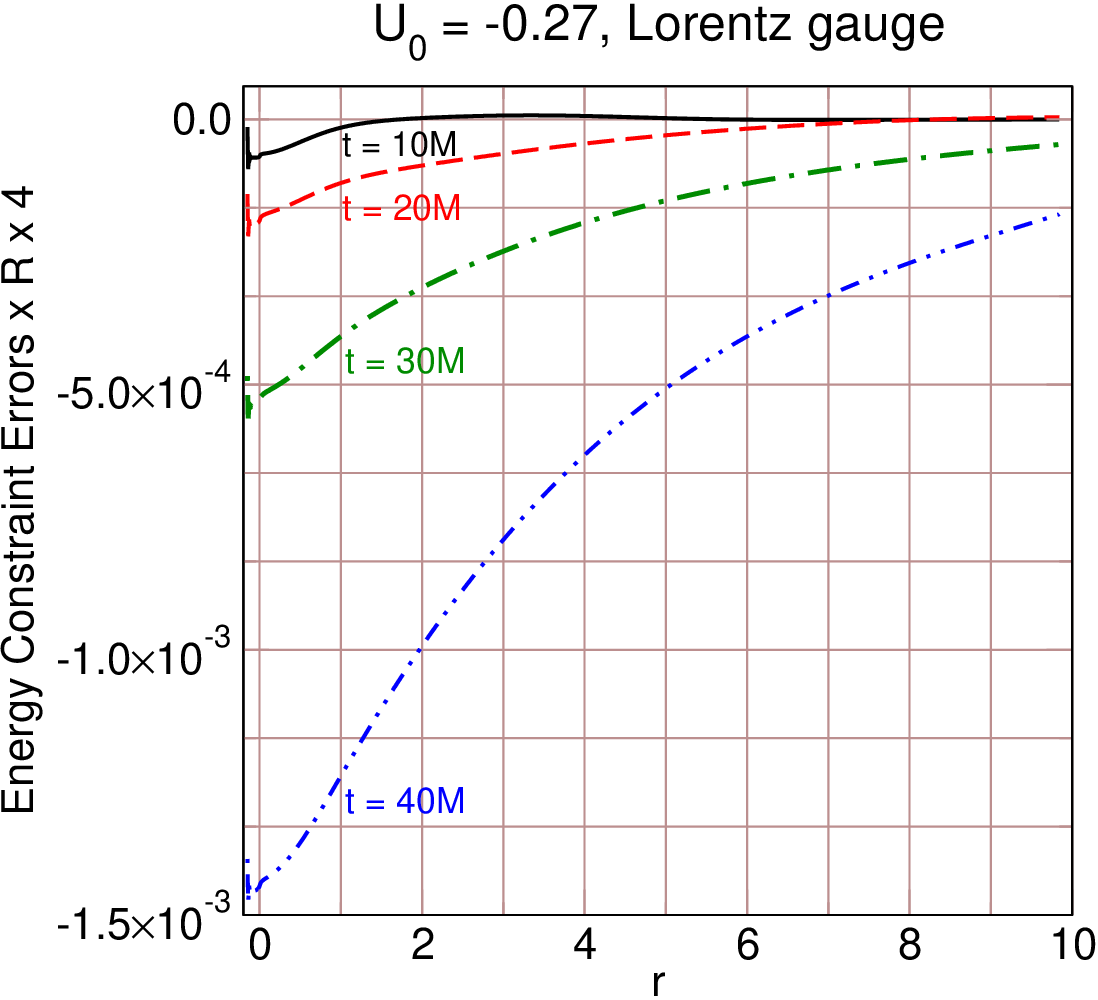}}
\caption{Constraint errors vs. time for stationary
IVP.  $dr=0.01$, $dt=0.001$.  The lapse and shift are reset
every $dt=0.002$.}
\vspace{2cm}
\label{Constraint_errors_vs_time}
\end{figure*}

To test the accuracy and stability of the time evolution with stationary  
initial conditions we can not only check how accurately the constraint 
equations of Sections \ref{trueconstraints} and \ref{Bconstraints} are 
satisfied, but also take advantage of the fact that any change in any of
the variables with time can only be due to some error in the calculation.  
The growth of some the constraint errors with time for both Nester and 
Lorentz gauges is shown in Fig. \ref{Constraint_errors_vs_time}.  The mass 
constraint error is the fractional difference between the mass as calculated 
at each grid point and each time from Eq. (\ref{firstintegral}) and the mass 
used in constructing the initial conditions ($M=1/2$ for our choice of 
units).  The energy constraint error is the difference of the left and 
right-hand sides of Eq. (\ref{trueenergy}), 
with the derivative of $n_{\hat{r}}$ 
evaluated numerically.  The momentum constraint error (not plotted) from 
Eq. (\ref{truemom}) is similar in magnitude to the energy constraint error.  
The energy constraint errors are 
weighted by a factor of $R$, which serves to make the errors more 
representative of fractional errors in individual terms of the energy 
constraint and also to make the amplitudes of the curves more uniform with 
radius.  We multiply the actual errors by a factor of four because
when testing convergence, we compare these curves with those from runs with double 
the point spacing.  Assuming the quadratic convergence expected from our 
second-order accurate numerical method, the errors quadruple when the spacing
between grid points is doubled.

\begin{figure*}[h]
\resizebox{2.3in}{2.3in}
{\includegraphics{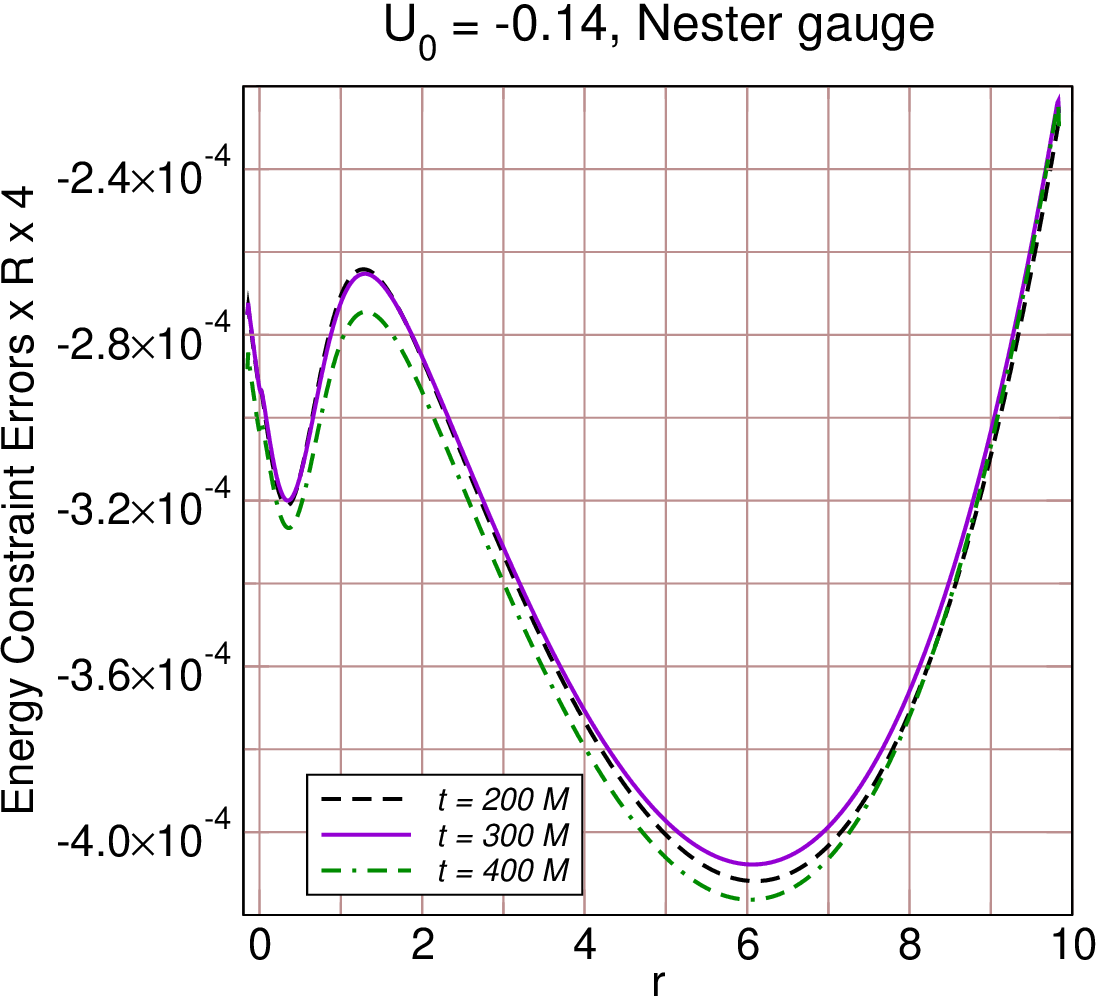}}
\caption{Energy constraint error at late times for 
stationary IVP and Nester gauge.  $dr=0.01$, $dt=0.001$. The lapse and 
shift are reset every $dt=0.002$. }
\vspace{2cm}
\label{EC_error_late_times}
\end{figure*}

The first two of the three plots show the results for the Nester gauge with 
$U_0=-0.14$, up to a time of $120~M$.  The errors do grow somewhat with time,
but the rate of growth, after an initial oscillation in the mass constraint 
error, is clearly decreasing.  The errors stop growing almost completely at 
later times, as shown in Fig. \ref{EC_error_late_times} for the energy 
constraint errors up to $t=400~M$.  The Lorentz gauge results for the energy 
constraint errors, the third plot in Fig. \ref{Constraint_errors_vs_time},
clearly show exponential growth with time.  The growth rate is not sensitive 
to numerical errors, and seems to be due to an analytically unstable 
constraint-violating mode of the evolution equations.

\begin{figure*}[h]
\resizebox{2.3in}{2.3in}{\includegraphics{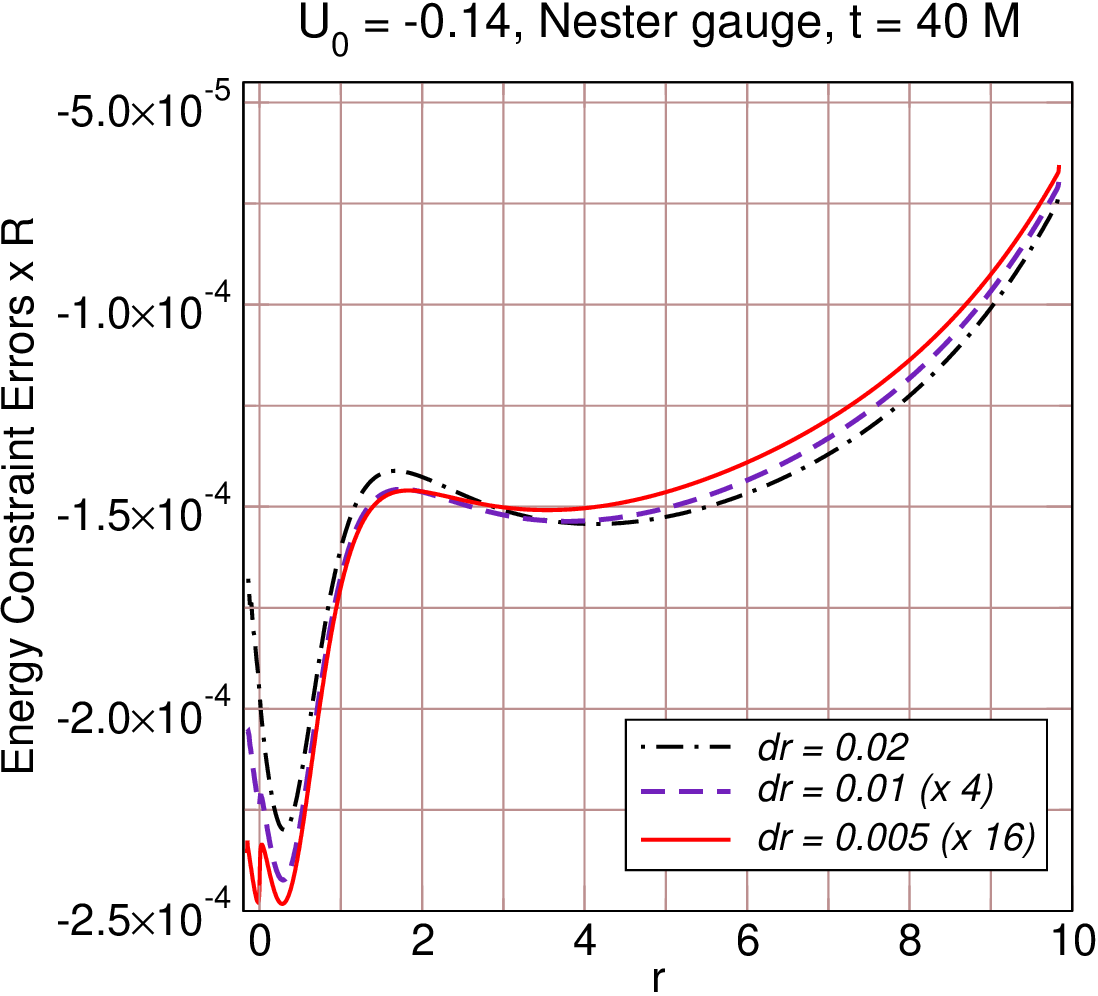}}
\caption{Convergence of energy constraint 
errors at $t = 40~M$ for stationary IVP and Nester gauge. 
The results at resolution $dr=0.01$ have been multiplied
by $4$, and those at $dr=0.005$ have been multiplied by $16$.  $dr/dt=10.0$ for
all resolutions.  The lapse and shift are reset every $dt = 0.002$.}
\vspace{2cm}
\label{EC_error_convergence}
\end{figure*}

The apparent stability of the time evolution for the Nester gauge is 
encouraging, but it is also important to demonstrate convergence of the 
errors as the grid spacing is reduced.  Fig. \ref{EC_error_convergence} 
does this for the Nester gauge evolution at $t=40~M$.  We compare results 
for $dr=0.02$, $dr=0.01$, and $dr=0.005$.  The errors are multiplied by 
the appropriate factors to make the curves lie on top of each other 
assuming perfect second-order convergence.  While there are some deviations 
close to the inner edge of the grid, the convergence is quite good 
everywhere else.  At later times the convergence does gradually deteriorate,
but that is not surprising after a very large number of dynamical times.

The good results for the Nester gauge with $U_0=-0.14$ 
unfortunately deteriorate 
rather rapidly as $U_0$ is varied away from this value.  For $U_0=-0.19$, for 
instance, the errors grow much more rapidly initially, and then start 
oscillating with slowly increasing amplitude for $t>100~M$ or so.  Convergence 
starts breaking down around $t=60M$.  The errors at late times are at least 
a couple of order of magnitudes greater than they are for $U_0=-0.14$.  There 
is no real indication of exponential growth, but accuracy is not very good 
even for $1000$ points in the grid.

The apparent need to fine-tune the initial conditions to get errors under good 
control with the Nester gauge suggests that the present equation formulation 
does not give the kind of robust accuracy and stability that is necessary 
for dealing with physically more interesting problems of black hole dynamics 
and mergers.  The Lorentz gauge, with its exponentially growing instabilities, 
seems clearly unsuitable for dealing with black hole problems.

\subsection{Constant Mean Curvature (CMC) IVP}

A more demanding test of the code is to start from initial 
conditions which do not correspond to a stationary solution, and 
see if the gauge conditions allow relaxation toward a stationary 
solution.  Certainly one only expects to be able to evolve 
black hole spacetimes accurately for many dynamical times if 
this is the case.  The obvious first choice, consistent with
the conformal approach to the initial value problem pioneered by
York \cite{BowenYork80,York99,PfeifferYork2003}, 
is to require that the trace of the extrinsic 
curvature be uniform on the initial hypersurface, which is what 
we mean by constant mean curvature.  This is the
condition for separability of the momentum and energy constraint 
equations in the conformal approach, and has been assumed in 
much of the work on initial value problem over the last few 
decades.  The most common assumption has been that the initial 
hypersurface is maximal, $\text{trace}~ K=0$, but hyperbolic 
hypersurfaces, with $\text{trace}~ K>0$, are more desirable from our point 
of view, since they simplify boundary conditions at the outer edge 
of the grid and seem to be necessary for the stability of our 
evolution equations.

The value of the constant mean curvature, $K_0$, is a free parameter,
as is the value of $U_0$ ($R~K_T$ on the horizon).  While the initial 
acceleration of the tetrad congruence is in principle a free 
function, we fix it according to the stationary condition of 
Eq. (\ref{aralg}).  This gives the time evolution a quiet start, in that
the initial partial time derivative of $R~K_T$ is zero, but nothing more
than that.

\begin{figure*}[h]
\resizebox{2in}{2.1in}{\includegraphics{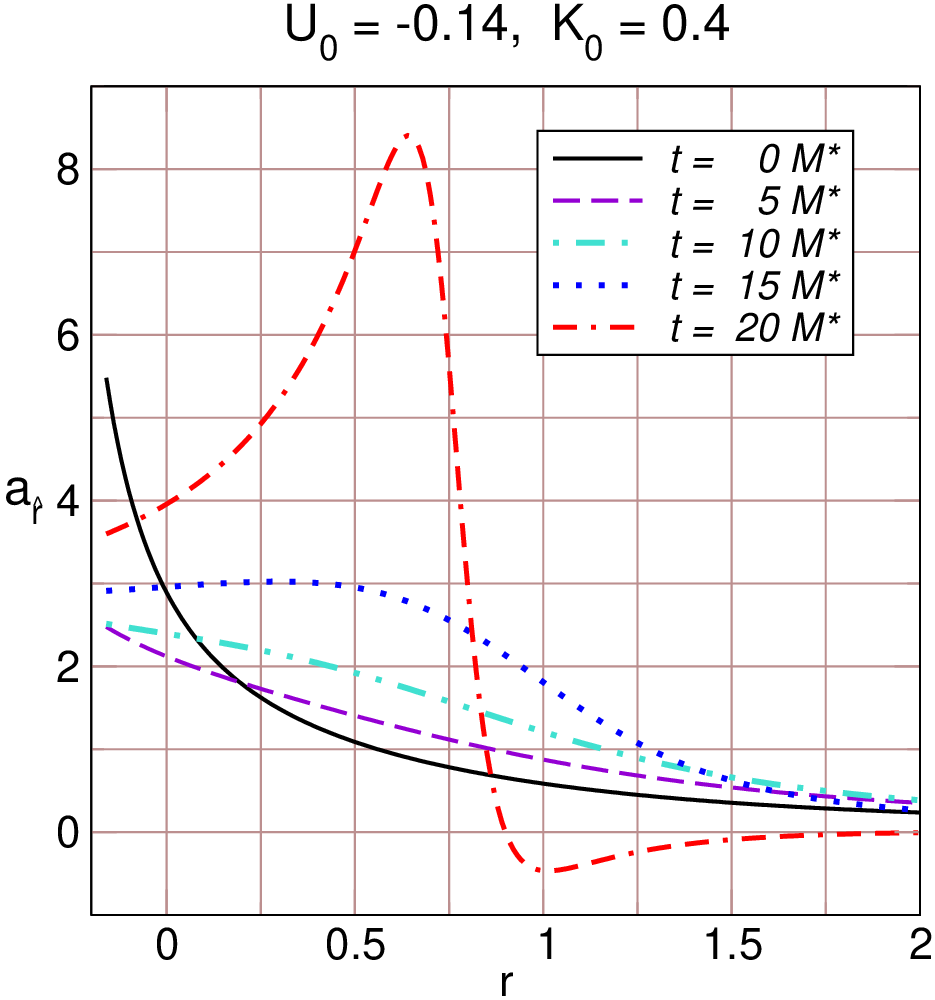}}
\hspace{.2in}
\resizebox{2in}{2.1in}{\includegraphics{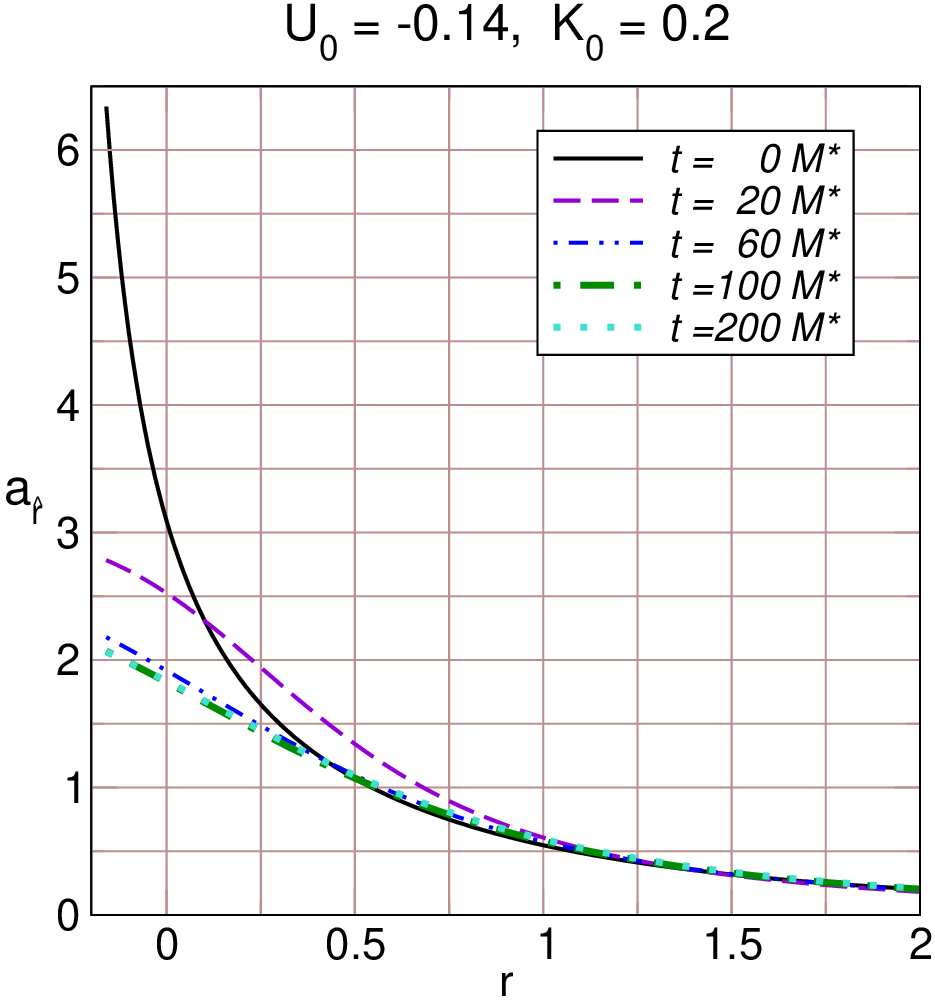}}
\hspace{.2in}
\resizebox{2in}{2.1in}{\includegraphics{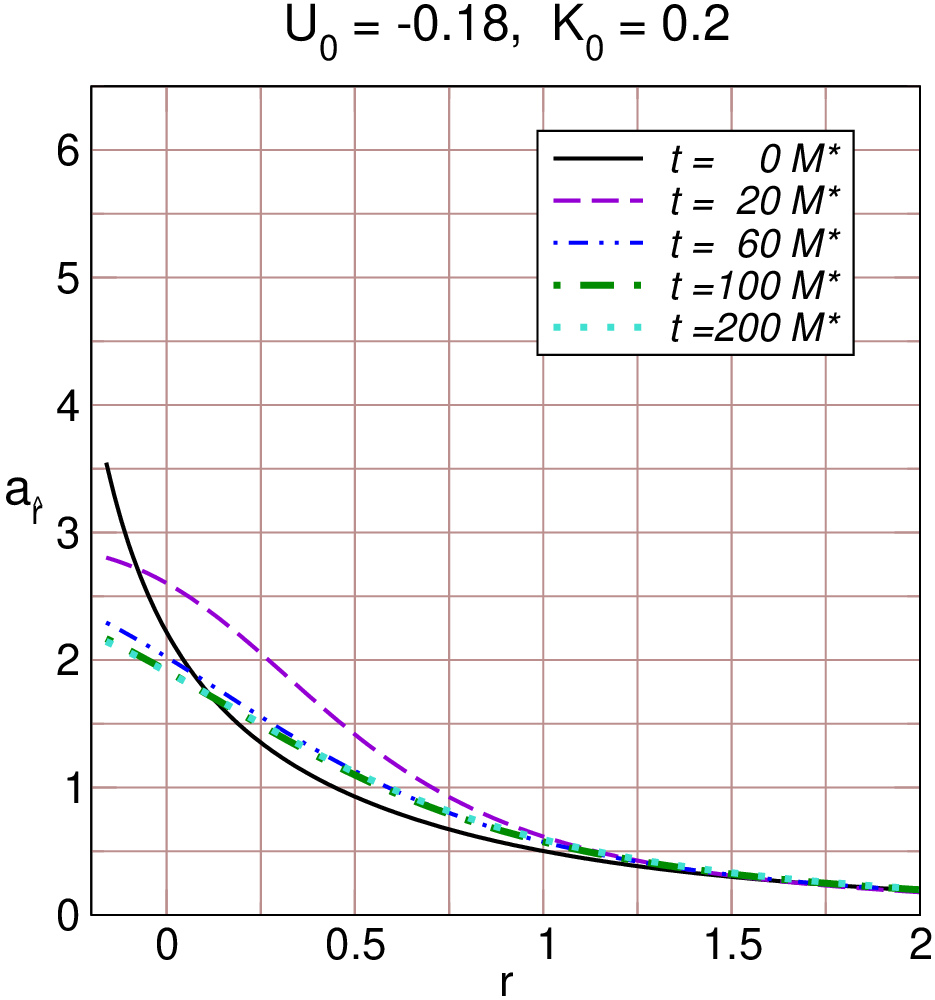}}
\caption{CMC initial conditions and Nester gauge:  
With $K_0=0.4$, $a_{\hat{r}}$ 
evolves pathologically, but with $K_0=0.2$, $a_{\hat{r}}$
approaches a stationary solution (shown is region near event horizon).  
*These times are for $r= 9.84$.}
\vspace{2cm}
\label{CMC_N_evolution}
\end{figure*}

Fig. \ref{CMC_N_evolution}
shows the Nester gauge time evolution of $a_{\hat{r}}$ for three different
choices of $K_0$ and $U_0$.
With $K_0=0.2$ and $U_0=-0.14$, the solution approaches a
stationary solution, with the value of $R~K_T$ on the event
horizon equal to $-0.15$ at $t=200~M$.  With $K_0=0.2$ and 
$U_0=-0.18$, the solution also approaches a stationary solution, 
with the value of $R~K_T$ on the horizon equal to $-0.16$ at $t=200~M$.   
However, with $K_0=0.4$ and $U_0=-0.14$, the solution becomes pathological
due to an increasingly sharp gradient of the acceleration somewhat outside 
the horizon, and the code crashes.

Similar initial conditions with the Lorentz gauge do not lead to any 
pathologies in the acceleration, but in all cases roughly exponential 
growth of constraint errors prevents meaningful continuation of the solutions 
much beyond a time of $40~M$.

\section{Discussion}

The WEBB tetrad formulation
for numerical relativity has been implemented and
tested in spherically symmetric black hole
spacetimes, with two types of tetrad gauge conditions, which we call 
the Nester and Lorentz gauge conditions.  While there is freedom in 
the choice of the initial velocity and acceleration of the tetrad 
congruence, subsequent evolution of the tetrad frames follows 
uniquely from the tetrad gauge conditions.  The coordinate 
evolution is determined by 
elliptic equations for a tetrad lapse and shift in such a way 
that hypersurfaces of constant time stay perpendicular to the 
tetrad congruence, and the spatial coordinates evolve according to 
a minimal deformation condition, with boundary conditions which 
keep both the inner and outer edges of the grid at roughly constant
Schwarzschild curvature radius $R$.  While we pay particular attention 
to initial conditions consistent with a stationary evolution, we
also consider non-stationary evolution from constant mean curvature
initial hypersurfaces.  
Some representative results have been presented, but we 
have experimented with a variety of other parameter choices, numerical methods,
etc.  We have done sufficient testing to convince ourselves that the 
instabilities we find are not due to problems with numerical methods, but 
represent genuine analytic instabilities of the evolution equations,
with numerical errors only playing a role of seeding the instabilities.

There are several potential factors which can influence the existence 
and severity of instabilities, such as (1) the choice of tetrad gauge, 
(2) the equation formulation, including choices of variables and the 
possible addition of constraint equations to the evolution equations in 
various combinations, (3) initial conditions and how the coordinates 
are evolved, and (4) boundary conditions.  Of course, all of these factors
interact in various ways and it is not always clear just what is 
behind a particular instability.

The equation formulation, gauge conditions, and the choice of 
variables determine the hyperbolic structure of the evolution system.  The 
WEBB framework makes specific choices for all three of these, which give 
a particularly simple symmetrizable hyperbolic system with all propagation 
either along the light cone or along the tetrad congruence.  All we do 
in this paper is adapt the variables to explicit spherical symmetry, and
compare two alternative tetrad gauge conditions, the Nester and Lorentz 
gauges.  The fairly trivial choices of variables that we have 
explored ($B_T \equiv 1/R$ versus $R$ and $B_R \equiv e^{-\lambda}$ versus
$\lambda$) have some effect on accuracy, but no real effect on stability.
 
We have discovered potential problems with both Nester and Lorentz gauges 
which cannot be fixed by playing with the equation formulation.  The Lorentz 
gauge is not consistent with stationary solutions of the evolution equations 
which have asymptotically hyperbolic spacelike hypersurfaces.  With the 
Nester gauge the tetrad congruence can develop a kind of gauge shock, 
with steep gradients in the acceleration, for certain initial conditions.  
These issues would remain even if other problems could be fixed with a change 
in equation formulation.

In experiments with various boundary conditions on the evolution 
equations, we found that it is critically important to use some form of
constraint-preserving boundary conditions.  Otherwise constraint errors 
are generated at the boundary and propagate into the grid at the speed 
of light.  The particular scheme we implement is not as elegant as some that
have been proposed in the literature, but seems to work well, at least in 
this simple one-dimensional context.  The other key to 
boundary conditions is to arrange to have the "zero-velocity" 
modes along the tetrad congruence propagate out of the grid at the boundary.  
This is no problem at the inner excision boundary, 
but requires expansion of the tetrad congruence 
at the outer boundary.  Since we keep our constant-time hypersurfaces 
close to being orthogonal to the tetrad congruence, this is equivalent 
to having asymptotically hyperbolic hypersurfaces bending toward the 
future.  With our choice of sign for extrinsic curvature, this means
positive extrinsic curvature for the hypersurface.

For the Nester gauge, there is only a rather narrow range of initial 
conditions which give rise to stable and reasonably accurate solutions
of the evolution equations.  For stationary initial 
conditions, if the parameter $U_0=R~K_T$ is too close to zero, 
the acceleration and the radial extrinsic curvature $K_R$ 
have very large gradients near the horizon, which is bad for stability.
However, $U_0$ needs to be much less negative than $-0.25$ in order that
the hyperbolic character of the hypersurface be strong enough
to give stability.  There is a rather narrow window around $U_0=-0.14$ where
constraint and other errors do not grow substantially.  We have not 
found any initial conditions for which the evolution in the Lorentz gauge
is stable.

The importance of positive extrinsic curvature for stability has also 
been noted in a much more rigorous analytic treatment of the stability of the
Weyl tensor evolution equations by Frauendiener and Vogel 
\cite{FrauendienerVogel04}.

Finally, we note that in the context of black hole evolution it is critical to 
control the relation of the inner edge of the coordinate grid to the apparent
event horizon, and highly desirable to keep the outer edge of the grid at a 
constant physical radius.  We do this through the boundary conditions on our 
minimal deformation equation for the shift vector.  
Failure to do this leads to severe numerical problems.

We conclude that the WEBB equation formulation as presented in
\cite{BuBa03}, with either the
Nester or Lorentz gauge conditions, does not seem particularly
promising for black hole evolutions.  While the Nester gauge can give good
results for some initial conditions, these results are not very robust.  
The Lorentz gauge does not seem 
to be viable at all.  However, it may be possible, by adding constraint
damping terms to the evolution equations, to improve accuracy and stability
for a wider range of initial conditions, while preserving a reasonable
hyperbolic structure.  We have begun to explore other equation formulations 
and gauge conditions in the context of the tetrad framework, some of which 
seem considerably more promising.

\begin{acknowledgments}
We gratefully acknowledge F. Estabrook and H. Wahlquist
for their insights and contributions, and
O. Sarbach for reading the manuscript.  
We thank K. Thorne and L. Lindblom for generously 
accommodating us during CalTech's numerical relativity visitors program.
L.T.B. was supported by the NASA Graduate Student Researchers Program
under Grant No. NGT5-50298 and by the National Research Council
Research Associateship Program under Contract No. NASW 99027.
\end{acknowledgments}

\bibliography{WEBB_Schwarz}

\begin{thebibliography}{45}
\expandafter\ifx\csname natexlab\endcsname\relax\def\natexlab#1{#1}\fi
\expandafter\ifx\csname bibnamefont\endcsname\relax
  \def\bibnamefont#1{#1}\fi
\expandafter\ifx\csname bibfnamefont\endcsname\relax
  \def\bibfnamefont#1{#1}\fi
\expandafter\ifx\csname citenamefont\endcsname\relax
  \def\citenamefont#1{#1}\fi
\expandafter\ifx\csname url\endcsname\relax
  \def\url#1{\texttt{#1}}\fi
\expandafter\ifx\csname urlprefix\endcsname\relax\def\urlprefix{URL }\fi
\providecommand{\bibinfo}[2]{#2}
\providecommand{\eprint}[2][]{\url{#2}}

\bibitem[{\citenamefont{M{\o}ller}(1961)}]{M61}
\bibinfo{author}{\bibfnamefont{C.}~\bibnamefont{M{\o}ller}},
  \bibinfo{journal}{Kgl. Danske Videnskab. Selskab, Mat-Fys. Skr.}
  \textbf{\bibinfo{volume}{1}}, \bibinfo{pages}{1} (\bibinfo{year}{1961}).

\bibitem[{\citenamefont{Newman and Penrose}(1962)}]{newman62}
\bibinfo{author}{\bibfnamefont{E.~T.} \bibnamefont{Newman}} \bibnamefont{and}
  \bibinfo{author}{\bibfnamefont{R.}~\bibnamefont{Penrose}},
  \bibinfo{journal}{J. Math. Phys.} \textbf{\bibinfo{volume}{3}},
  \bibinfo{pages}{566} (\bibinfo{year}{1962}), \bibinfo{note}{erratum {\bf 4}
  998(E) (1963)}.

\bibitem[{\citenamefont{Estabrook and Wahlquist}(1964)}]{EW64}
\bibinfo{author}{\bibfnamefont{F.~B.} \bibnamefont{Estabrook}}
  \bibnamefont{and} \bibinfo{author}{\bibfnamefont{H.~D.}
  \bibnamefont{Wahlquist}}, \bibinfo{journal}{J. Math. Phys.}
  \textbf{\bibinfo{volume}{5}}, \bibinfo{pages}{1629} (\bibinfo{year}{1964}).

\bibitem[{\citenamefont{Ashtekar}(1986)}]{AA86}
\bibinfo{author}{\bibfnamefont{A.}~\bibnamefont{Ashtekar}},
  \bibinfo{journal}{Phys. Rev. Lett.} \textbf{\bibinfo{volume}{57}},
  \bibinfo{pages}{2244} (\bibinfo{year}{1986}).

\bibitem[{\citenamefont{Ashtekar}(1987)}]{AA87}
\bibinfo{author}{\bibfnamefont{A.}~\bibnamefont{Ashtekar}},
  \bibinfo{journal}{Phys. Rev. D} \textbf{\bibinfo{volume}{36}},
  \bibinfo{pages}{1587} (\bibinfo{year}{1987}).

\bibitem[{\citenamefont{Friedrich}(1996)}]{HF96}
\bibinfo{author}{\bibfnamefont{H.}~\bibnamefont{Friedrich}},
  \bibinfo{journal}{Class. Quantum Grav.} \textbf{\bibinfo{volume}{13}},
  \bibinfo{pages}{1451} (\bibinfo{year}{1996}).

\bibitem[{\citenamefont{van Putten and Eardley}(1996)}]{VPE96}
\bibinfo{author}{\bibfnamefont{M.~H. P.~M.} \bibnamefont{van Putten}}
  \bibnamefont{and} \bibinfo{author}{\bibfnamefont{D.~M.}
  \bibnamefont{Eardley}}, \bibinfo{journal}{Phys. Rev. D}
  \textbf{\bibinfo{volume}{53}}, \bibinfo{pages}{3056} (\bibinfo{year}{1996}).

\bibitem[{\citenamefont{Estabrook et~al.}(1997)\citenamefont{Estabrook,
  Robinson, and Wahlquist}}]{ERW97}
\bibinfo{author}{\bibfnamefont{F.~B.} \bibnamefont{Estabrook}},
  \bibinfo{author}{\bibfnamefont{R.~S.} \bibnamefont{Robinson}},
  \bibnamefont{and} \bibinfo{author}{\bibfnamefont{H.~D.}
  \bibnamefont{Wahlquist}}, \bibinfo{journal}{Class. Quantum Grav.}
  \textbf{\bibinfo{volume}{14}}, \bibinfo{pages}{1237} (\bibinfo{year}{1997}).

\bibitem[{\citenamefont{van Elst and Uggla}(1997)}]{VEU97}
\bibinfo{author}{\bibfnamefont{H.}~\bibnamefont{van Elst}} \bibnamefont{and}
  \bibinfo{author}{\bibfnamefont{C.}~\bibnamefont{Uggla}},
  \bibinfo{journal}{Class. Quantum Grav.} \textbf{\bibinfo{volume}{14}},
  \bibinfo{pages}{2673} (\bibinfo{year}{1997}).

\bibitem[{\citenamefont{Iriondo et~al.}(1998)\citenamefont{Iriondo,
  Leguizam{\'o}n, and Reula}}]{ILR98}
\bibinfo{author}{\bibfnamefont{M.~S.} \bibnamefont{Iriondo}},
  \bibinfo{author}{\bibfnamefont{E.~O.} \bibnamefont{Leguizam{\'o}n}},
  \bibnamefont{and} \bibinfo{author}{\bibfnamefont{O.~A.} \bibnamefont{Reula}},
  \bibinfo{journal}{Adv. Theor. Math. Phys.} \textbf{\bibinfo{volume}{2}},
  \bibinfo{pages}{1075} (\bibinfo{year}{1998}).

\bibitem[{\citenamefont{Yoneda and Shinkai}(1999)}]{YS99}
\bibinfo{author}{\bibfnamefont{G.}~\bibnamefont{Yoneda}} \bibnamefont{and}
  \bibinfo{author}{\bibfnamefont{H.}~\bibnamefont{Shinkai}},
  \bibinfo{journal}{Phys. Rev. Lett.} \textbf{\bibinfo{volume}{82}},
  \bibinfo{pages}{263} (\bibinfo{year}{1999}).

\bibitem[{\citenamefont{Yoneda and Shinkai}(2000)}]{YS00}
\bibinfo{author}{\bibfnamefont{G.}~\bibnamefont{Yoneda}} \bibnamefont{and}
  \bibinfo{author}{\bibfnamefont{H.}~\bibnamefont{Shinkai}},
  \bibinfo{journal}{Int. J. Mod. Phys. D} \textbf{\bibinfo{volume}{9}},
  \bibinfo{pages}{13} (\bibinfo{year}{2000}).

\bibitem[{\citenamefont{Jantzen et~al.}(2001)\citenamefont{Jantzen, Carini, and
  Bini}}]{Jantzen01}
\bibinfo{author}{\bibfnamefont{R.~T.} \bibnamefont{Jantzen}},
  \bibinfo{author}{\bibfnamefont{P.}~\bibnamefont{Carini}}, \bibnamefont{and}
  \bibinfo{author}{\bibfnamefont{D.}~\bibnamefont{Bini}},
  \emph{\bibinfo{title}{Understanding spacetime splittings and their
  relationships or gravitoelectromagnetism: the user manual}}
  (\bibinfo{year}{2001}),
  \bibinfo{note}{http://www34.homepage.villanova.edu/robert.jantzen/gem/}.

\bibitem[{\citenamefont{Choquet-Bruhat and York}()}]{CBY02}
\bibinfo{author}{\bibfnamefont{Y.}~\bibnamefont{Choquet-Bruhat}}
  \bibnamefont{and} \bibinfo{author}{\bibfnamefont{J.~W.} \bibnamefont{York}},
  \bibinfo{note}{gr-qc/0202014}.

\bibitem[{\citenamefont{Buchman and Bardeen}(2003)}]{BuBa03}
\bibinfo{author}{\bibfnamefont{L.~T.} \bibnamefont{Buchman}} \bibnamefont{and}
  \bibinfo{author}{\bibfnamefont{J.~M.} \bibnamefont{Bardeen}},
  \bibinfo{journal}{Phys. Rev. D} \textbf{\bibinfo{volume}{67}},
  \bibinfo{pages}{084017} (\bibinfo{year}{2003}), \bibinfo{note}{erratum {\bf
  72} 049903(E) (2005)}.

\bibitem[{\citenamefont{Frauendiener}(2004)}]{Frauendiener04}
\bibinfo{author}{\bibfnamefont{J.}~\bibnamefont{Frauendiener}},
  \bibinfo{journal}{Living Rev. Relativity} \textbf{\bibinfo{volume}{7}},
  \bibinfo{pages}{1} (\bibinfo{year}{2004}), \bibinfo{note}{(cited on 8/26/05):
  http://www.livingreviews.org/lrr-2004-1}.

\bibitem[{\citenamefont{Bardeen}()}]{Bardeen05}
\bibinfo{author}{\bibfnamefont{J.~M.} \bibnamefont{Bardeen}}, \bibinfo{note}{in
  preparation}.

\bibitem[{\citenamefont{Estabrook}(2005{\natexlab{a}})}]{Estabrook05a}
\bibinfo{author}{\bibfnamefont{F.~B.} \bibnamefont{Estabrook}},
  \bibinfo{journal}{Phys. Rev. D} \textbf{\bibinfo{volume}{71}},
  \bibinfo{pages}{044004} (\bibinfo{year}{2005}{\natexlab{a}}).

\bibitem[{\citenamefont{Estabrook}(2005{\natexlab{b}})}]{Estabrook05b}
\bibinfo{author}{\bibfnamefont{F.~B.} \bibnamefont{Estabrook}}
  (\bibinfo{year}{2005}{\natexlab{b}}), \bibinfo{note}{gr-qc/0508081}.

\bibitem[{\citenamefont{Nester}(1992)}]{JN92}
\bibinfo{author}{\bibfnamefont{J.~M.} \bibnamefont{Nester}},
  \bibinfo{journal}{J. Math. Phys.} \textbf{\bibinfo{volume}{33}},
  \bibinfo{pages}{910} (\bibinfo{year}{1992}).

\bibitem[{\citenamefont{Smarr and York}(1978)}]{SY78}
\bibinfo{author}{\bibfnamefont{L.}~\bibnamefont{Smarr}} \bibnamefont{and}
  \bibinfo{author}{\bibfnamefont{J.~W.} \bibnamefont{York}},
  \bibinfo{journal}{Phys. Rev. D} \textbf{\bibinfo{volume}{17}},
  \bibinfo{pages}{2529} (\bibinfo{year}{1978}).

\bibitem[{\citenamefont{Stewart}(1998)}]{Stewart98}
\bibinfo{author}{\bibfnamefont{J.~M.} \bibnamefont{Stewart}},
  \bibinfo{journal}{Class. Quantum Grav.} \textbf{\bibinfo{volume}{15}},
  \bibinfo{pages}{2865} (\bibinfo{year}{1998}).

\bibitem[{\citenamefont{Friedrich and Nagy}(1999)}]{FN99}
\bibinfo{author}{\bibfnamefont{H.}~\bibnamefont{Friedrich}} \bibnamefont{and}
  \bibinfo{author}{\bibfnamefont{G.}~\bibnamefont{Nagy}},
  \bibinfo{journal}{Commun. Math. Phys.} \textbf{\bibinfo{volume}{201}},
  \bibinfo{pages}{619} (\bibinfo{year}{1999}).

\bibitem[{\citenamefont{Iriondo and Reula}(2002)}]{IR02}
\bibinfo{author}{\bibfnamefont{M.~S.} \bibnamefont{Iriondo}} \bibnamefont{and}
  \bibinfo{author}{\bibfnamefont{O.~A.} \bibnamefont{Reula}},
  \bibinfo{journal}{Phys. Rev. D} \textbf{\bibinfo{volume}{65}},
  \bibinfo{pages}{044024} (\bibinfo{year}{2002}).

\bibitem[{\citenamefont{Bardeen and Buchman}(2002)}]{BaBu02}
\bibinfo{author}{\bibfnamefont{J.~M.} \bibnamefont{Bardeen}} \bibnamefont{and}
  \bibinfo{author}{\bibfnamefont{L.~T.} \bibnamefont{Buchman}},
  \bibinfo{journal}{Phys. Rev. D} \textbf{\bibinfo{volume}{65}},
  \bibinfo{pages}{064037} (\bibinfo{year}{2002}).

\bibitem[{\citenamefont{Szilagyi and Winicour}(2003)}]{SW03}
\bibinfo{author}{\bibfnamefont{B.}~\bibnamefont{Szilagyi}} \bibnamefont{and}
  \bibinfo{author}{\bibfnamefont{J.}~\bibnamefont{Winicour}},
  \bibinfo{journal}{Phys. Rev. D} \textbf{\bibinfo{volume}{68}},
  \bibinfo{pages}{041501} (\bibinfo{year}{2003}).

\bibitem[{\citenamefont{Calabrese et~al.}(2003)\citenamefont{Calabrese, Pullin,
  Reula, Sarbach, and Tiglio}}]{CPRST03}
\bibinfo{author}{\bibfnamefont{G.}~\bibnamefont{Calabrese}},
  \bibinfo{author}{\bibfnamefont{J.}~\bibnamefont{Pullin}},
  \bibinfo{author}{\bibfnamefont{O.}~\bibnamefont{Reula}},
  \bibinfo{author}{\bibfnamefont{O.}~\bibnamefont{Sarbach}}, \bibnamefont{and}
  \bibinfo{author}{\bibfnamefont{M.}~\bibnamefont{Tiglio}},
  \bibinfo{journal}{Commun. Math. Phys.} \textbf{\bibinfo{volume}{240}},
  \bibinfo{pages}{377} (\bibinfo{year}{2003}).

\bibitem[{\citenamefont{Szil\'{a}gyi et~al.}(2002)\citenamefont{Szil\'{a}gyi,
  Schmidt, and Winicour}}]{SSW02}
\bibinfo{author}{\bibfnamefont{B.}~\bibnamefont{Szil\'{a}gyi}},
  \bibinfo{author}{\bibfnamefont{B.}~\bibnamefont{Schmidt}}, \bibnamefont{and}
  \bibinfo{author}{\bibfnamefont{J.}~\bibnamefont{Winicour}},
  \bibinfo{journal}{Phys. Rev. D} \textbf{\bibinfo{volume}{65}},
  \bibinfo{pages}{064015} (\bibinfo{year}{2002}).

\bibitem[{\citenamefont{Calabrese et~al.}(2002)\citenamefont{Calabrese, Lehner,
  and Tiglio}}]{CLT02}
\bibinfo{author}{\bibfnamefont{G.}~\bibnamefont{Calabrese}},
  \bibinfo{author}{\bibfnamefont{L.}~\bibnamefont{Lehner}}, \bibnamefont{and}
  \bibinfo{author}{\bibfnamefont{M.}~\bibnamefont{Tiglio}},
  \bibinfo{journal}{Phys. Rev. D} \textbf{\bibinfo{volume}{65}},
  \bibinfo{pages}{104031} (\bibinfo{year}{2002}).

\bibitem[{\citenamefont{Calabrese and Sarbach}(2003)}]{CS03}
\bibinfo{author}{\bibfnamefont{G.}~\bibnamefont{Calabrese}} \bibnamefont{and}
  \bibinfo{author}{\bibfnamefont{O.}~\bibnamefont{Sarbach}},
  \bibinfo{journal}{J. Math. Phys.} \textbf{\bibinfo{volume}{44}},
  \bibinfo{pages}{3888} (\bibinfo{year}{2003}).

\bibitem[{\citenamefont{Frittelli and G\'{o}mez}(2004)}]{frittelligomez04}
\bibinfo{author}{\bibfnamefont{S.}~\bibnamefont{Frittelli}} \bibnamefont{and}
  \bibinfo{author}{\bibfnamefont{R.}~\bibnamefont{G\'{o}mez}},
  \bibinfo{journal}{Phys. Rev. D} \textbf{\bibinfo{volume}{69}},
  \bibinfo{pages}{124020} (\bibinfo{year}{2004}).

\bibitem[{\citenamefont{Lindblom et~al.}(2004)\citenamefont{Lindblom, Scheel,
  Kidder, Pfeiffer, Shoemaker, and Teukolsky}}]{LSKPST04}
\bibinfo{author}{\bibfnamefont{L.}~\bibnamefont{Lindblom}},
  \bibinfo{author}{\bibfnamefont{M.~A.} \bibnamefont{Scheel}},
  \bibinfo{author}{\bibfnamefont{L.~E.} \bibnamefont{Kidder}},
  \bibinfo{author}{\bibfnamefont{H.~P.} \bibnamefont{Pfeiffer}},
  \bibinfo{author}{\bibfnamefont{D.}~\bibnamefont{Shoemaker}},
  \bibnamefont{and} \bibinfo{author}{\bibfnamefont{S.~A.}
  \bibnamefont{Teukolsky}}, \bibinfo{journal}{Phys. Rev. D}
  \textbf{\bibinfo{volume}{69}}, \bibinfo{pages}{124025}
  (\bibinfo{year}{2004}).

\bibitem[{\citenamefont{Kidder et~al.}(2005)\citenamefont{Kidder, Lindblom,
  Scheel, Buchman, and Pfeiffer}}]{KLSBP05}
\bibinfo{author}{\bibfnamefont{L.~E.} \bibnamefont{Kidder}},
  \bibinfo{author}{\bibfnamefont{L.}~\bibnamefont{Lindblom}},
  \bibinfo{author}{\bibfnamefont{M.~A.} \bibnamefont{Scheel}},
  \bibinfo{author}{\bibfnamefont{L.~T.} \bibnamefont{Buchman}},
  \bibnamefont{and} \bibinfo{author}{\bibfnamefont{H.~P.}
  \bibnamefont{Pfeiffer}}, \bibinfo{journal}{Phys. Rev. D}
  \textbf{\bibinfo{volume}{71}}, \bibinfo{pages}{064020}
  (\bibinfo{year}{2005}).

\bibitem[{\citenamefont{Sarbach and Tiglio}(2004)}]{ST05}
\bibinfo{author}{\bibfnamefont{O.}~\bibnamefont{Sarbach}} \bibnamefont{and}
  \bibinfo{author}{\bibfnamefont{M.}~\bibnamefont{Tiglio}}
  (\bibinfo{year}{2004}), \bibinfo{note}{gr-qc/0412115; Journal of Hyperbolic
  Differential Equations (in press)}.

\bibitem[{\citenamefont{van Putten}(1997)}]{VP97}
\bibinfo{author}{\bibfnamefont{M.~H. P.~M.} \bibnamefont{van Putten}},
  \bibinfo{journal}{Phys. Rev. D} \textbf{\bibinfo{volume}{55}},
  \bibinfo{pages}{4705} (\bibinfo{year}{1997}).

\bibitem[{\citenamefont{Frauendiener}(1998)}]{Frauendiener98}
\bibinfo{author}{\bibfnamefont{J.}~\bibnamefont{Frauendiener}},
  \bibinfo{journal}{Phys. Rev. D} \textbf{\bibinfo{volume}{58}},
  \bibinfo{pages}{064003} (\bibinfo{year}{1998}).

\bibitem[{\citenamefont{Shinkai and Yoneda}(2000)}]{SY00}
\bibinfo{author}{\bibfnamefont{H.}~\bibnamefont{Shinkai}} \bibnamefont{and}
  \bibinfo{author}{\bibfnamefont{G.}~\bibnamefont{Yoneda}},
  \bibinfo{journal}{Class. Quantum Grav.} \textbf{\bibinfo{volume}{17}},
  \bibinfo{pages}{4799} (\bibinfo{year}{2000}).

\bibitem[{\citenamefont{Yoneda and Shinkai}(2001)}]{YS01}
\bibinfo{author}{\bibfnamefont{G.}~\bibnamefont{Yoneda}} \bibnamefont{and}
  \bibinfo{author}{\bibfnamefont{H.}~\bibnamefont{Shinkai}},
  \bibinfo{journal}{Class. Quantum Grav.} \textbf{\bibinfo{volume}{18}},
  \bibinfo{pages}{441} (\bibinfo{year}{2001}).

\bibitem[{\citenamefont{Garfinkle}(2004)}]{DF04}
\bibinfo{author}{\bibfnamefont{D.}~\bibnamefont{Garfinkle}},
  \bibinfo{journal}{Phys. Rev. Lett.} \textbf{\bibinfo{volume}{93}},
  \bibinfo{pages}{161101} (\bibinfo{year}{2004}).

\bibitem[{\citenamefont{Bowen and {York, Jr}}(1980)}]{BowenYork80}
\bibinfo{author}{\bibfnamefont{J.~M.} \bibnamefont{Bowen}} \bibnamefont{and}
  \bibinfo{author}{\bibfnamefont{J.~W.} \bibnamefont{{York, Jr}}},
  \bibinfo{journal}{Phys. Rev. D} \textbf{\bibinfo{volume}{21}},
  \bibinfo{pages}{2047} (\bibinfo{year}{1980}).

\bibitem[{\citenamefont{{York, Jr}}(1999)}]{York99}
\bibinfo{author}{\bibfnamefont{J.~W.} \bibnamefont{{York, Jr}}},
  \bibinfo{journal}{Phys. Rev. Lett.} \textbf{\bibinfo{volume}{82}},
  \bibinfo{pages}{1350} (\bibinfo{year}{1999}).

\bibitem[{\citenamefont{Pfeiffer and {York, Jr.}}(2003)}]{PfeifferYork2003}
\bibinfo{author}{\bibfnamefont{H.~P.} \bibnamefont{Pfeiffer}} \bibnamefont{and}
  \bibinfo{author}{\bibfnamefont{J.~W.} \bibnamefont{{York, Jr.}}},
  \bibinfo{journal}{Phys. Rev. D} \textbf{\bibinfo{volume}{67}},
  \bibinfo{pages}{044022} (\bibinfo{year}{2003}).

\bibitem[{\citenamefont{Reula}(1998)}]{Reula98}
\bibinfo{author}{\bibfnamefont{O.~A.} \bibnamefont{Reula}},
  \bibinfo{journal}{Living Rev. Relativity} \textbf{\bibinfo{volume}{1}},
  \bibinfo{pages}{3} (\bibinfo{year}{1998}), \bibinfo{note}{(cited on 8/26/05):
  http://www.livingreviews.org/lrr-1998-3}.

\bibitem[{\citenamefont{Press et~al.}(1986)\citenamefont{Press, Flannery,
  Teukolsky, and Vetterling}}]{Press86}
\bibinfo{author}{\bibfnamefont{W.~H.} \bibnamefont{Press}},
  \bibinfo{author}{\bibfnamefont{B.~P.} \bibnamefont{Flannery}},
  \bibinfo{author}{\bibfnamefont{S.~A.} \bibnamefont{Teukolsky}},
  \bibnamefont{and} \bibinfo{author}{\bibfnamefont{W.~T.}
  \bibnamefont{Vetterling}}, \emph{\bibinfo{title}{Numerical Recipes}}
  (\bibinfo{publisher}{Cambridge University Press},
  \bibinfo{address}{Cambridge, England}, \bibinfo{year}{1986}).

\bibitem[{\citenamefont{Frauendiener and Vogel}(2005)}]{FrauendienerVogel04}
\bibinfo{author}{\bibfnamefont{J.}~\bibnamefont{Frauendiener}}
  \bibnamefont{and} \bibinfo{author}{\bibfnamefont{T.}~\bibnamefont{Vogel}},
  \bibinfo{journal}{Class. Quant. Grav.} \textbf{\bibinfo{volume}{22}},
  \bibinfo{pages}{1769} (\bibinfo{year}{2005}).

\end{thebibliography}
 
\end{document}